\documentclass[12pt, draftclsnofoot, onecolumn]{IEEEtran}
\usepackage{amsfonts}
\usepackage{amsmath}
\usepackage{amssymb}
\usepackage{bm}
\usepackage{xcolor,graphicx,float}
\usepackage[font=small,justification=centering]{caption}
\usepackage{diagbox}
\usepackage{color, soul}
\usepackage{stfloats}
\usepackage[numbers,sort&compress]{natbib}
\usepackage[amsmath,thmmarks]{ntheorem}
\usepackage{theorem} 
\usepackage{algorithm}
\usepackage{algorithmic}
\usepackage{graphicx}
\usepackage{subfigure}
\usepackage{pict2e}

\theoremheaderfont{\sc}\theorembodyfont{\upshape}
\theoremstyle{nonumberplain}
\theoremsymbol{\rule{1ex}{1ex}}

\begin{document}
\title{Expectation Propagation based Line Spectral Estimation}
\author{Jiang~Zhu, Xupeng Lei, Mihai Alin-Badiu and Fengzhong Qu}
\maketitle
\begin{abstract}
The fundamental problem of line spectral estimation (LSE) using the expectation propagation (EP) method is studied. Previous approaches estimate the model order sequentially, limiting their practical utility in scenarios with large dimensions of measurements and signals. To overcome this limitation, a bilinear generalized LSE (BiG-LSE) method that concurrently estimates the model order is developed. The key concept involves iteratively approximating the original nonlinear model as a bilinear model through Taylor series expansion, with EP employed for inference. To mitigate computational complexity, the posterior log-pdfs are approximated to reduce the number of messages. BiG-LSE automatically determines the model order, noise variance, provides uncertainty levels for the estimates, and adeptly handles nonlinear measurements. Based on the BiG-LSE, a variant employing the von Mises distribution for the frequency is developed, which is suitable for sequential estimation.
Numerical experiments and real data are used to demonstrate that BiG-LSE achieves estimation accuracy comparable to current methods.
\end{abstract}
\section{Introduction}
Line spectral estimation (LSE), which involves extracting frequencies and amplitudes from a noisy mixture of sinusoids, has garnered considerable attention \cite{Stoica}. This problem is encountered in various engineering, physics, and natural science disciplines, including radar signal processing, channel estimation, and direction of arrival (DOA) estimation.


Compressed sensing (CS) methods are employed to tackle LSE under the assumption that the number of frequencies is small compared to the number of measurements. By constraining frequencies to a grid, the LSE problem is transformed into a sparse reconstruction problem, often referred to as on-grid methods. However, as frequencies are continuous-valued and do not precisely align with the grid, this can result in frequency leakage and lead to model mismatch issues \cite{ChiMis}. To address these issues, off-grid and grid-less CS methods, such as atomic norm minimization (ANM)-based methods \cite{GLS}, optimization-based methods \cite{FangIRA}, greedy-based methods \cite{Mamandipoor}, and Bayesian methods \cite{Hu1Newtonref, Hu2Taylorexpan, Badiu, HanseSuperfast}, have been proposed.

ANM-based methods are theoretically guaranteed to recover frequencies in the noise-free scenario under a minimum separation condition \cite{GLS}. However, they entail solving a semidefinite programming (SDP) problem, the computational complexity of which becomes prohibitively large even for moderately sized problems. The optimization-based methods presented in \cite{FangIRA} utilize an iterative reweighted approach, refining frequencies in parallel through a Newton approach, resulting in high computational complexity. The notable greedy-based method, Newtonalized orthogonal matching pursuit (NOMP), performs LSE sequentially \cite{Mamandipoor}. Typically, several Newton steps are needed to achieve estimation accuracy close to the CRB. The computation complexity of NOMP scales with $O(KN\log(N)+NK^3+K^4)$, where $K$ and $N$ denote the number of frequencies and measurements, respectively. Consequently, as the number of frequencies $K$ increases, the computation complexity of NOMP is expected to rise significantly.

In \cite{Hu1Newtonref, Hu2Taylorexpan, Badiu, HanseSuperfast}, a Bayesian perspective is adopted for the LSE problem. In \cite{Hu1Newtonref}, a hierarchical Gaussian prior is imposed on the coefficient vector in sparse Bayesian learning to induce sparsity. The Newton method is employed to refine frequencies, but the computational complexity, scaling with $O(N^3)$, is exceptionally high. Building on the work in \cite{zhuTLS} and \cite{Yangoff}, \cite{Hu2Taylorexpan} utilizes the first-order approximation of the true dictionary to address basis mismatch. It jointly estimates amplitudes and grid bias. The computation complexity of this method scales with $O(qN_g\log N_g)$, with $q$ and $N_g \gg N$ representing the number of iterations required by conjugate gradient (CG) and the number of grids, where $q$ can be as large as $N_g$. In \cite{Badiu}, a Bernoulli-Gaussian distribution is introduced to model complex weights conditioned on activation variables. Treating frequencies as random variables, the variational LSE (VALSE) algorithm is proposed \cite{Badiu}. This method automatically estimates noise variance, nuisance parameters, determines the model order, and provides uncertain degrees of frequency estimates, contrasting with previous works that offer point estimates. This approach is natural for sequential estimation. The computation complexity of VALSE is $O(N^2 + NK^3)$, which is also high for large dimensions of measurements or signals. To alleviate this, a superfast LSE inspired by VALSE is introduced, achieving estimation accuracy at least comparable to current methods but operating orders of magnitude faster \cite{HanseSuperfast}. The computation complexity of superfast LSE is $O(N\log(N) + NK^2 + K^3)$. It is evident that when the number of frequencies $K$ is large, such as $K \approx \rho N$ where $\rho$ is a constant, the aforementioned methods become computationally intensive and time-consuming.

The approximate message passing (AMP) algorithm, serving as an approximation of the sum-product belief propagation algorithm, proves to be an efficient inference method for linear inverse problems \cite{AMP}. Later, AMP is extended to generalized AMP (GAMP) \cite{GAMP} to address the generalized linear model (GLM), which can be derived via expectation propagation (EP) \cite{mengunified, Minka}. GAMP performs well when the elements of the measurement matrix are drawn from a subGaussian distribution or when the matrix is a partial discrete Fourier transform (DFT) matrix. However, it may diverge otherwise. As a result, GAMP can be directly applied to solve LSE with a computation complexity of $O(N\log N)$ due to the DFT matrix structure, but it suffers from model mismatch issues \cite{ChiMis}. 

In \cite{BilinearGAMP1, BilinearGAMP2}, the bilinear GAMP (BiG-AMP) algorithm is derived as an approximation of the sum-product belief propagation algorithm. BiG-AMP finds applications in matrix completion, robust principal component analysis (PCA), and dictionary learning. It is demonstrated that BiG-AMP achieves excellent reconstruction accuracy with a computation complexity of $O(MNL)$, where the dimensions of the measurement matrix and signal are  ${\mathbb R}^{M\times N}$ and ${\mathbb R}^{N\times L}$, respectively. Importantly, the computation involves only matrix multiplication and can be implemented in parallel, making it hardware-friendly. To address the correlated measurement matrix scenario,  unitary AMP (UAMP) \cite{BiUAMP} based on unitary transformation on the measurement matrix are proposed, demonstrating robustness. 

In this study, motivated by the low computational complexity of BiG-AMP and with the aim of addressing high dimensional LSE involving multiple frequencies, we propose a Bayesian approach called bilinear generalized LSE (BiG-LSE). This method is also capable of handling LSE problems arising from nonlinear measurements. The nonlinearity in this problem manifests in two ways: first, the LSE model depends nonlinearly on frequencies, and second, the measurements themselves may be nonlinear. To address the first issue, we perform a Taylor series expansion at fixed known frequencies on the original nonlinear model, thereby transforming it into a bilinear model. To tackle the second issue, the EP method is well-suited for handling nonlinear measurements by introducing component-wise nonlinear minimum mean squared error (MMSE) or maximum a posteriori (MAP) modules \cite{mengunified}. In deriving the BiG-LSE algorithm, EP is employed, and the number of messages is reduced through approximation.

Unlike BiG-AMP, BiG-LSE requires the estimation of parameters involving both real-valued frequencies and complex-valued amplitudes, whereas BiG-AMP estimates either real or complex parameters alone. Additionally, the fast Fourier transform (FFT) is incorporated to accelerate the BiG-LSE algorithm. Since the true frequencies are close to the fixed known frequencies, the Taylor series expansion yields accurate results. As a result, oversampling is employed to obtain dense grids. However, the use of dense grids introduces high correlation in the dictionary, potentially causing BiG-LSE to diverge. To enhance robustness and estimation accuracy, we iteratively apply BiG-LSE by constructing a dimension-reduced dictionary, see the BiG-LSE software \cite{BiGLSEsoftware}. BiG-LSE automatically estimates the model order, noise variance, provides uncertainty levels for the estimates, and effectively manages nonlinear measurements.

Our conference version \cite{EPLSE} was the first to explore LSE using EP, where the derivation is omitted. However, it directly utilized messages computed by EP to estimate frequencies without simplifying them, leading to high computational complexity. As an extended version, we present the detailed derivation, streamline the process by discarding insignificant terms and reducing the number of messages. Furthermore, while the von Mises distribution was used to model the prior of frequencies in EPLSE, the BiG-LSE framework divides the frequency space into uniformly spaced grids and employs a Gaussian distribution to model the bias between the true frequency and its nearest frequency grid point. The BiG-LSE is also extended to incorporate the von Mises prior of the frequency. Numerical and real experiments demonstrate high estimation accuracy compared to state-of-the-art methods. In conclusion, this work makes a significant contribution to enhancing the practical viability of the AMP framework.

\section{Problem Setup and Probabilistic Model}
Let ${\mathbf z}_{\rm LS}\in {\mathbb C}^N$ be a line spectral (LS) signal consisting of $K$ complex sinusoids
\begin{align}\label{signal-generate}
{\mathbf z}_{\rm LS}=\sum\limits_{k=1}^K  {\mathbf a}({\theta}_k){x}_k={\mathbf A}({\boldsymbol\theta}){\mathbf x},
\end{align}
where ${x}_k$ is the complex amplitude of the $k$th frequency, ${\theta}_k\in [0,2\pi)$ is the $k$th frequency, ${\mathbf A}({\boldsymbol\theta})=[{\mathbf a}({\theta}_1),{\mathbf a}({\theta}_2),\cdots,{\mathbf a}({\theta}_K)]^{\rm T}$ and
\begin{align}
{\mathbf a}(\theta)=[1,{\rm e}^{{\rm j}\theta},\cdots,{\rm e}^{{\rm j}(N-1)\theta}]^{\rm T}/\sqrt{N}.
\end{align}
The LS ${\mathbf z}_{\rm LS}$ undergoes a componentwise (linear or nonlinear) transform. For simplicity, we focus on the scenario such that the LS is corrupted by the additive white Gaussian noise (AWGN) described by
\begin{align}\label{meas}
	{\mathbf y}={\mathbf z}_{\rm LS}+{\mathbf w}_{\rm LS},
\end{align}
where ${\mathbf w}_{\rm LS}\sim {{\mathcal {CN}}({\mathbf w}_{\rm LS};{\mathbf 0},\sigma_{\rm LS}^2{\mathbf I}_N)}$, $\sigma^2_{\rm LS}$ is the variance of the noise. The proposed algorithm that we have made available also addresses the nonlinear measurement scenario ${\mathbf y}={\mathcal Q}_{\rm C}({\mathbf z}_{\rm LS}+{\mathbf w}_{\rm LS})$, where ${\mathcal Q}_{\rm C}({\mathbf x})\triangleq {\mathcal Q}(\Re\{{\mathbf x}\}\})+{\rm j}{\mathcal Q}(\Im\{{\mathbf x}\})$, ${\mathcal Q}(\cdot)$ denotes nonlinear operation such as a low resolution quantizer. Note that either the model (\ref{meas}) or the nonlinear measurement model can be described by a conditional PDF $p({\mathbf y}|{\mathbf z}_{\rm LS};\sigma_{\rm LS}^2)$ given as
\begin{align}
	p({\mathbf y}|{\mathbf z}_{\rm LS};\sigma_{\rm LS}^2)=\prod\limits_{n=1}^Np(y_n|[{\mathbf z}_{\rm LS}]_n;\sigma_{\rm LS}^2).
\end{align}
For the $k$th frequency, we define its SNR as 
\begin{align}
	{\rm SNR}_k={\|{\mathbf a}({\theta}_k){x}_k\|_2^2}/{\sigma^2}={|{x}_k|_2^2}/(N\sigma^2).
\end{align}
Note that the above SNRs correspond to the sample SNR. The integrated SNR is $10\log N$ dB above the sample SNR.

Because the frequencies ${\boldsymbol\theta}$ are nonlinearly coupled with the LS ${\mathbf z}_{\rm LS}$, it is hard to perform efficient estimation. Let $a_{nk}({\theta}_{k})$ denote the $(n,k)$th element of ${\mathbf A}({\boldsymbol\theta})$ defined in (\ref{signal-generate}). We perform Taylor series expansion
on $a_{nk}(\theta_k)$ at $\bar{\theta}_{k}$ near $\theta_k$, it produces 
\begin{align}\label{zLsTaylor}
	[{\mathbf z}_{\rm LS}]_n  = \underbrace{\sum\limits_{k=1}^K (a_{nk}(\bar{\theta}_{k})+b_{nk}(\bar{\theta}_{k})\epsilon_k)x_k}_{z_n}+\underbrace{\sum\limits_{k=1}^KO(\epsilon_k^2)}_{[{\boldsymbol\epsilon}_{\rm res}]_n},
\end{align}
where 
\begin{subequations}
	\begin{align}
		b_{nk}(\bar{\theta})&\triangleq \frac{1}{N}\frac{\partial a_{nk}(\theta)}{\partial \theta}\big|_{\theta=\bar{\theta}},\label{bnkdef}\\
		\epsilon_k&\triangleq N(\theta_k-\bar{\theta}_{k}).\label{epsilonnkdef}
	\end{align}
\end{subequations}
Since it is difficult to estimate ${\boldsymbol\epsilon}_{\rm res}$, we treat it as the noise and we define the equivalent noise $\mathbf w$ as 
\begin{align}\label{eqnoisew}
	{\mathbf w}\triangleq {\mathbf w}_{\rm LS}+{\boldsymbol\epsilon}_{\rm res}.
\end{align}
For simplicity, we assume that  ${\mathbf w}\sim {{\mathcal {CN}}({\mathbf w};{\mathbf 0},\sigma^2{\mathbf I}_N)}$, $\sigma^2$ is the variance of the noise ${\mathbf w}$. Define $\bar{\boldsymbol \theta}=[\bar{\theta}_{1},\bar{\theta}_{2},\cdots,\bar{\theta}_{K}]^{\rm T}$, ${\boldsymbol\epsilon}=[\epsilon_1,\epsilon_2,\cdots,\epsilon_K]^{\rm T}$. $\mathbf z$ can be written in matrix-vector form as 
\begin{align}\label{bilinearz}
	\mathbf{z} = \left({\mathbf A}(\bar{\boldsymbol \theta}) + {\mathbf B}(\bar{\boldsymbol \theta}){\rm diag}({\boldsymbol \epsilon})\right){\mathbf x}.
\end{align}
According to (\ref{signal-generate}), (\ref{meas}), (\ref{eqnoisew}) and (\ref{bilinearz}), we obtain the following matrix-vector models
	\begin{align}
		{\mathbf y}&={\mathbf z}+{\mathbf w},\label{AdditiveBilmodel}
	\end{align}
where $\mathbf{A}(\bar{\boldsymbol{\theta}}) = [\mathbf{a}(\bar{\theta}_{1}),\cdots,\mathbf{a}(\bar{\theta}_{K})]$, $\mathbf{B}(\bar{\boldsymbol{\theta}}) = [\partial \mathbf{a}(\bar{\theta}_{1})/\partial \bar{\theta}_{1},\partial \mathbf{a}(\bar{\theta}_{2})/\partial \bar{\theta}_{2},\cdots,\partial \mathbf{a}(\bar{\theta}_{K})/\partial \bar{\theta}_{K}]/N$. 
Besides, the more closer the $\bar{\theta}_k$ to the true frequency $\theta_k$, the more accurate the model approximation becomes, resulting in a smaller energy of ${\boldsymbol\epsilon}_{\rm res}$. 
Since the sparsity level $K$ is usually unknown, the line spectral signal consisting of $L$ complex sinusoids is assumed \cite{Badiu}. Define  
\begin{subequations}
	\begin{align}
\bar{\boldsymbol \theta}=[\bar{\theta}_{1},\bar{\theta}_{2},\cdots,\bar{\theta}_{L}]^{\rm T},\label{thetabar}\\
{\boldsymbol\epsilon}=[\epsilon_1,\epsilon_2,\cdots,\epsilon_L]^{\rm T},\label{epsilonvecdef}
\end{align}
\end{subequations}
where $\bar{\boldsymbol \theta}$ is supposed to be known. The approach to obtain $\bar{\boldsymbol \theta}$ will be deferred in Subsection \ref{Initsubsec}. According to (\ref{AdditiveBilmodel}), we obtain 
\begin{align}
	\mathbf{z} = \left({\mathbf A}(\bar{\boldsymbol \theta}) + {\mathbf B}(\bar{\boldsymbol \theta}){\rm diag}({\boldsymbol \epsilon})\right){\mathbf x},
\end{align}
and the following overcomplete matrix-vector model
	\begin{align}
		{\mathbf y}={\mathbf z}+{\mathbf w}, \label{sparselinmodel}
	\end{align}
where $\mathbf{A}(\bar{\boldsymbol{\theta}}) = [\mathbf{a}(\bar{\theta}_{1}),\cdots,\mathbf{a}(\bar{\theta}_{L})]$, $\mathbf{B}(\bar{\boldsymbol{\theta}}) = [\partial \mathbf{a}(\bar{\theta}_{1})/\partial \bar{\theta}_{1},\partial \mathbf{a}(\bar{\theta}_{2})/\partial \bar{\theta}_{2},\cdots,\partial \mathbf{a}(\bar{\theta}_{L})/\partial \bar{\theta}_{L}]/N$. Since the number of true spectral line is $K$ and $K\ll L$ as often the case, we construct an overcomplete measurement model.

Define $s_l$ as the indicator variable of the $l$th frequency, where $s_l=1$ denotes that the $l$th frequency is active, and $s_l=0$ denotes that the $l$th frequency is inactive. Let
\begin{align}
	p(s_l=1)=\pi_l, p(s_l=0)=1-\pi_l,
\end{align}
where $\pi_l$ is the activation probability of the $l$th frequency. Obviously, given $s_l$, the conditional distribution of $x_l$ is
\begin{align}\label{condxl}
	p(x_l|s_l;{\boldsymbol \omega}_x)=(1-s_l)\delta(x_l)+s_l{\mathcal{CN}}(x_l;\mu_0^x,\tau_0^x),
\end{align}
where $\mu_0$ and $\tau_0$ are the prior mean and variance of $x_l$  provided that $x_l$ is nonzero, ${\boldsymbol \omega}_x=[\pi_1,\cdots,\pi_L,\mu_0,\tau_0]^{\rm T}\in {\mathbb R}^{L+2}$ are the nuisance parameters involved in distribution of $p(x_l;{\boldsymbol \omega}_x)$. Here we impose the Gaussian distribution for $\epsilon_l$, i.e.,
\begin{align}\label{epsilondistribution}
	p(\epsilon_l)={\mathcal{N}}(\epsilon_l;0,\sigma_{\epsilon}^2),
\end{align}
where $\sigma_{\epsilon}^2$ will be determined later in Subsection \ref{Initsubsec}. It is worth noting that directly performing inference on the unknown parameters is intractable. As a result, an iterative algorithm is designed in Section \ref{EPLSE}.

\section{BiG-LSE Algorithm} \label{EPLSE}
\begin{figure*}[htbp]
    \centering
    \includegraphics[width=0.85\textwidth]{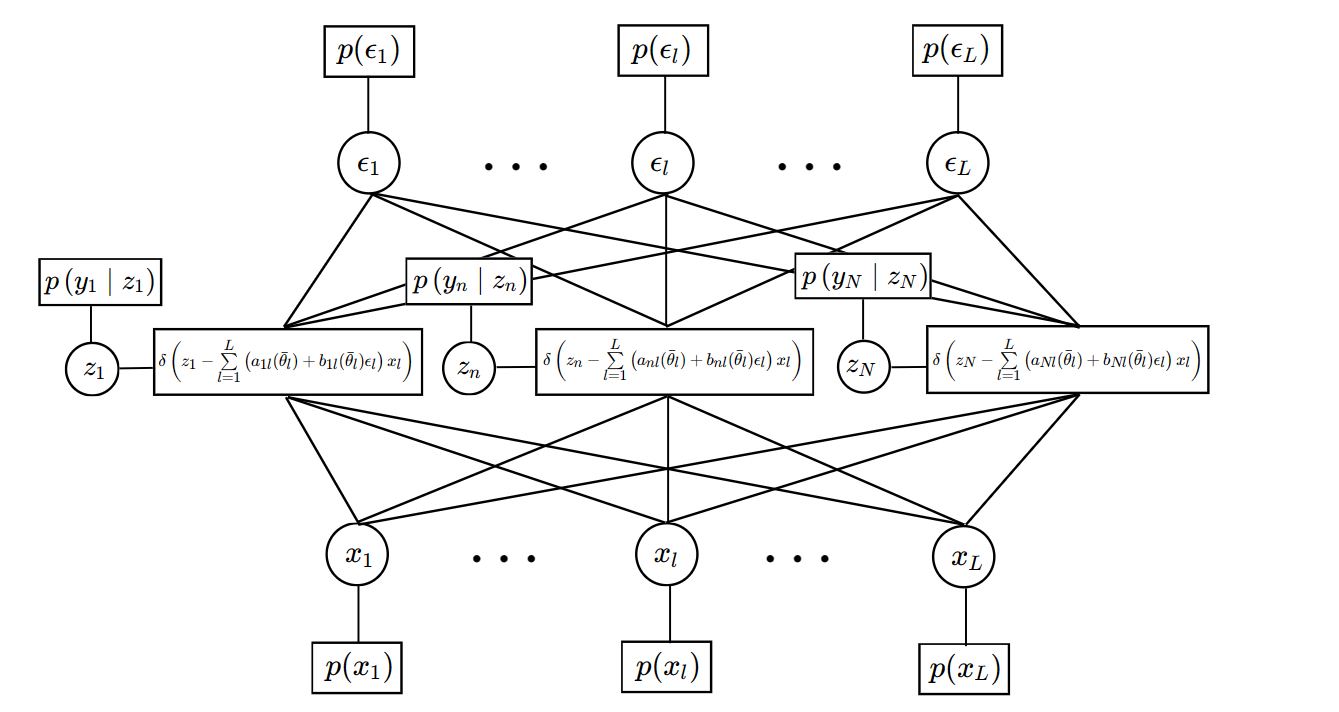}
    \caption{The whole factor graph. Here we introduce a delta factor node $\delta(\cdot)$, which simplifies the calculation as shown later .}
    \label{totalfac}
\end{figure*}

The factor graph is presented in Fig. \ref{totalfac} and a delta factor node $\delta(\cdot)$ is introduced. Before derivation, the following notations in Table \ref{Notations} are used.
First, we initialize $m_{\delta\rightarrow z}({\mathbf z})={\mathcal {CN}}({\mathbf z},{\mathbf z}_{\rm A}^{\rm ext},{\rm diag}({\mathbf v}_{\rm A}^{\rm ext}))$, ${m}_{l\rightarrow n}(x_l)\triangleq\mathcal{CN}(x_l;x_{l\rightarrow n},\nu^{x}_{l\rightarrow n})$ and $m_{l\rightarrow n}(\epsilon_l)\triangleq{\mathcal{N}}(\epsilon_l;\epsilon_{l\rightarrow n},\nu^\epsilon_{l\rightarrow n})$. Then we show how to update the messages to perform LSE.
\begin{table*}[h!t]
	\begin{center}
	\caption{Notations used for derivation}\label{Notations}
	\scalebox{0.95}{
        \begin{tabular}{|l|l|}
			\hline
			$m_{\delta\rightarrow z_n}(z_n)\triangleq{\mathcal{CN}}(z_n;{z}_{{\rm A},n}^{\rm ext},{v}_{{\rm A},n}^{\rm ext})$ & The message from the factor node $\delta\left(z_n-\sum\limits_{l=1}^L (a_{nl}(\bar{\theta}_{l})+b_{nl}(\bar{\theta}_{l})\epsilon_l)x_l\right)$ to the variable node $z_n$\\
			${m}_{z_n\rightarrow \delta}(z_n)\triangleq{\mathcal{CN}}(z_n;{z}_{{\rm B},n}^{\rm ext},{v}_{{\rm B},n}^{\rm ext})$ & The message from the variable node $z_n$ to the factor node $\delta\left(z_n-\sum\limits_{l=1}^L (a_{nl}(\bar{\theta}_{l})+b_{nl}(\bar{\theta}_{l})\epsilon_l)x_l\right)$\\
			$m_{n\rightarrow l}(\epsilon_l)\triangleq{\mathcal{N}}(\epsilon_l;\epsilon_{n\rightarrow l},\nu^{\epsilon}_{n\rightarrow l})$ & The message from the factor node $\delta\left(z_n-\sum\limits_{l=1}^L (a_{nl}(\bar{\theta}_{l})+b_{nl}(\bar{\theta}_{l})\epsilon_l)x_l\right)$ to the variable node $\epsilon_l$ \\
			$m_{n\rightarrow l}(x_l)\triangleq{\mathcal{CN}}(x_l;x_{n\rightarrow l},\nu^{x}_{n\rightarrow l})$ & The message from the factor node $\delta\left(z_n-\sum\limits_{l=1}^L (a_{nl}(\bar{\theta}_{l})+b_{nl}(\bar{\theta}_{l})\epsilon_l)x_l\right)$ to the variable node $x_l$ \\
			${m}_{l\rightarrow n}(\epsilon_l)\triangleq{\mathcal{N}}(\epsilon_l;\epsilon_{l\rightarrow n},\nu^{\epsilon}_{l\rightarrow n})$ & The message from the variable node $\epsilon_l$ to the factor node $\delta\left(z_n-\sum\limits_{l=1}^L (a_{nl}(\bar{\theta}_{l})+(b_{nl}\bar{\theta}_{l})\epsilon_l)x_l\right)$\\
			${m}_{l\rightarrow n}(x_l)\triangleq{\mathcal{CN}}(x_l;x_{l\rightarrow n},\nu^{x}_{l\rightarrow n})$ & The message from the variable node $x_l$ to the factor node $\delta\left(z_n-\sum\limits_{l=1}^L (a_{nl}(\bar{\theta}_{l})+b_{nl}(\bar{\theta}_{l})\epsilon_l)x_l\right)$ \\
			\hline
		\end{tabular}
        }
	\end{center}
\end{table*}
\subsection{Message Update}
\subsubsection{Update $m_{z\rightarrow \delta}({\mathbf z})$}\label{mzdelta}
\begin{figure}[ht]
	\centering
	\includegraphics[width=0.65\textwidth]{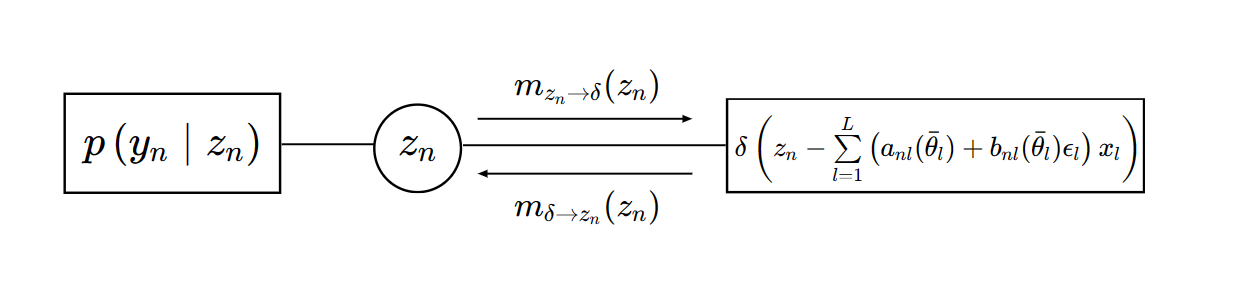}
	\caption{The subfactor graph for updating $m_{z_n\rightarrow \delta}(z_n)$. }
	\label{deltaz}
\end{figure}
According to EP, $m_{\delta \rightarrow z }({\mathbf z})={\mathcal {CN}}({\mathbf z},{\mathbf z}_{\rm A}^{\rm ext}(t),{\rm diag}({\mathbf v}_{\rm A}^{\rm ext}(t)))$ can be updated as
\begin{small}
   \begin{align}\label{extB0}
m_{z\rightarrow \delta}({\mathbf z})=\frac{{\rm Proj}\left[m_{\delta\rightarrow z}({\mathbf z})p({\mathbf y}|{\mathbf z};{\boldsymbol \omega}_z^{\rm old})\right]}{m_{\delta\rightarrow z}({\mathbf z})}\triangleq \frac{{\rm Proj}\left[q_{\rm B}(\mathbf z)\right]}{m_{\delta\rightarrow z}({\mathbf z})},
\end{align} 
\end{small}
where ${\rm Proj}\left[f({\mathbf z})\right]$ denotes the projection of the possibly nonnormalized probability density function (PDF) $f({\mathbf z})$ onto Gaussian PDF via moment matching. Then we calculate the componentwise posterior means and variances of $\mathbf z$ with respect to $q_{\rm B}(\mathbf z)$ as
\begin{align}
{\mathbf z}_{\rm B}^{\rm post}&={\rm E}\left[{\mathbf z}|q_{\rm B}(\mathbf z)\right],\\
{\mathbf v}_{\rm B}^{\rm post}&={\rm Var}\left[{\mathbf z}|q_{\rm B}(\mathbf z)\right].
\end{align}
Therefore ${\rm Proj}\left[q_{\rm B}(\mathbf z)\right]$ is 
\begin{align}\label{postz}
{\rm Proj}\left[q_{\rm B}(\mathbf z)\right]={\mathcal{CN}}({\mathbf z};{\mathbf z}_{\rm B}^{\rm post},{\rm diag}({\mathbf v}_{\rm B}^{\rm post})).
\end{align}
Note that for the linear measurement model ${\mathbf y}={\mathbf z}+{\mathbf w}$, we have ${\rm Proj}\left[q_{\rm B}(\mathbf z)\right]=q_{\rm B}(\mathbf z)$, i.e., the projection operation is the identify mapping. According to (\ref{extB0}), $m_{z\rightarrow \delta}({\mathbf z})$ is calculated to be
\begin{align}
m_{z\rightarrow \delta}({\mathbf z})={\mathcal{CN}}({\mathbf z};{\mathbf z}_{\rm B}^{\rm ext},{\rm diag}({\mathbf v}_{\rm B}^{\rm ext})),
\end{align}
where ${\mathbf v}_{\rm B}^{\rm ext}$ and ${\mathbf z}_{\rm B}^{\rm ext}$ are \cite{mengunified}
\begin{subequations}
	\begin{align}
	&{\mathbf v}_{\rm B}^{\rm ext}(t)=\left(\frac{1}{{\mathbf v}_{\rm B}^{\rm post}(t)}-\frac{1}{{\mathbf v}_{\rm A}^{\rm ext}(t)}\right)^{-1},\label{extB_var}\\
	&{\mathbf z}_{\rm B}^{\rm ext}(t)={\mathbf v}_{\rm B}^{\rm ext}(t)\left(\frac{{\mathbf z}_{\rm B}^{\rm post}(t)}{{\mathbf v}_{\rm B}^{\rm post}(t)}-\frac{{\mathbf z}_{\rm A}^{\rm ext}(t)}{{\mathbf v}_{\rm A}^{\rm ext}(t)}\right).\label{extB_mean}
	\end{align}
\end{subequations}

\subsubsection{Updating of $m_{n\rightarrow l}(\epsilon_l)$}\label{tildemthetan}
\begin{figure}[!ht]
	\centering
	\includegraphics[width=0.65\textwidth]{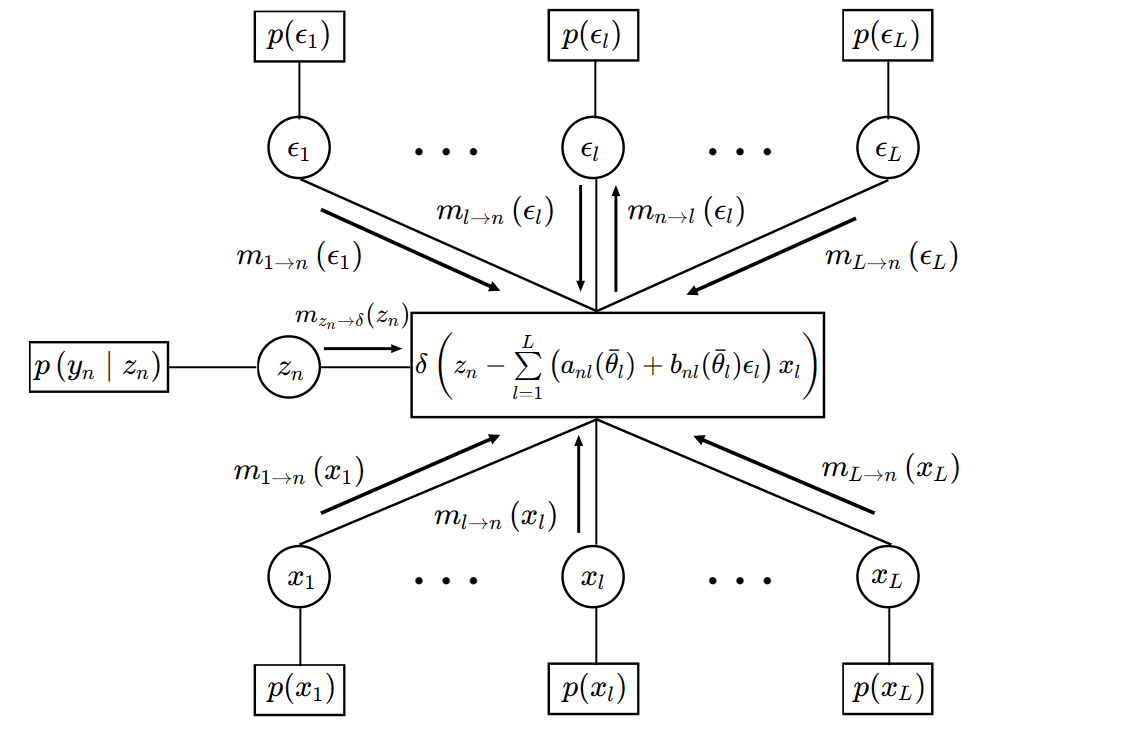}
	\caption{The factor graph for calculating $m_{n\rightarrow l}(\epsilon_l)$.}
	\label{equation11}
\end{figure}

The subfactor graph is presented in Fig. \ref{equation11}. 
According to EP, $m_{n\rightarrow l}(\epsilon_l)$ is updated as
\begin{align}
    &m_{n\rightarrow l}(\epsilon_l)\propto \notag\\
    &\frac{{\rm {Proj}}\left[\int\prod\limits_{l=1}^L\left({m}_{l\rightarrow n}(\epsilon_{l}){m}_{l\rightarrow n}(x_{l})\right)\delta\left(z_{n}-\sum\limits_{l^{\prime}=1}^L (a_{nl^{\prime}}(\bar{\theta}_{l^{\prime}})+b_{nl^{\prime}}(\bar{\theta}_{l^{\prime}})\epsilon_{l^{\prime}})x_{l^{\prime}}\right)m_{z_n\rightarrow\delta}(z_n)\prod\limits_{l=1}^L{\rm d}x_{l}\prod\limits_{l^{\prime}=1,l^{\prime}\neq l}^L{\rm d}\epsilon_{l^{\prime}}{\rm d}z_n\right]}{{m}_{l\rightarrow n}(\epsilon_l)}\notag\\
    &=\frac{{\rm {Proj}}\left[m_{l\rightarrow n}(\epsilon_l){\rm e}^{f_{n}(\epsilon_l)}\right]}{{m}_{l\rightarrow n}(\epsilon_l)} \approx {\rm Proj}\left[{\rm e}^{f_n(\epsilon_l)}\right]. \label{origtheta}.
\end{align}
Now we calculate the numerator term. First, $\delta\left(z_{n}-\sum\limits_{l^{\prime}=1}^L (a_{nl^{\prime}}(\bar{\theta}_{l^{\prime}})+b_{nl^{\prime}}(\bar{\theta}_{l^{\prime}})\epsilon_{l^{\prime}})x_{l^{\prime}}\right)$ and $m_{z_n\rightarrow\delta}(z_n)$ implies the following pseudo measurement model
\begin{small}
\begin{align}\label{defznm}
    \tilde{y}_{n}&=\sum\limits_{l^{\prime}=1}^L(a_{nl^{\prime}}(\bar{\theta}_{l^{\prime}})+b_{nl^{\prime}}(\bar{\theta}_{l^{\prime}})\epsilon_{l^{\prime}})x_{l^{\prime}}+\tilde{w}_n\notag\\
	&=\underbrace{(a_{nl}(\bar{\theta}_{l})+b_{nl}(\bar{\theta}_{l})\epsilon_{l})x_l+\left(\sum\limits_{l^{\prime}=1,l^{\prime}\neq l}^L(a_{nl^{\prime}}(\bar{\theta}_{l^{\prime}})+b_{nl^{\prime}}(\bar{\theta}_{l^{\prime}})\epsilon_{l^{\prime}})x_{l^{\prime}}\right)}_{z_n}+\tilde{w}_n,
\end{align}
\end{small}
where $\tilde{w}_n\sim {\mathcal{CN}}(\tilde{w}_n;0,\tilde{\sigma}_n^2)$, $\tilde{\sigma}_n^2={\mathbf v}_{{\rm B},n}^{\rm ext}$ and $\tilde{y}_{n}={\mathbf z}_{{\rm B},n}^{\rm ext}$ are the $n$th element of ${\mathbf v}_{\rm B}^{\rm ext}$ and ${\mathbf z}_{\rm B}^{\rm ext}$. Conditioned on $\epsilon_l$, we approximate $z_{n}$ as a Gaussian distribution by averaging over $\{\epsilon_{l^{\prime}}\}_{l^{\prime}=1,l^{\prime}\neq l}^L$ and $\{x_{l^{\prime}}\}_{l^{\prime}=1}^L$.
Utilizing $\Re\left\{a_{nl}(\bar{\theta}_l)b^*_{nl}(\bar{\theta}_l)\right\} = 0$, straightforward calculation yields
\begin{small}
    \begin{align}
    &{\rm E}\left[z_{n}|\epsilon_l\right]=\left((a_{nl}(\bar{\theta}_{l})+b_{nl}(\bar{\theta}_{l})\epsilon_{l})\right){\rm E}\left[x_l\right]+\sum\limits_{l^{\prime}=1,l^{\prime}\neq l}^L{\rm E}\left[(a_{nl^{\prime}}(\bar{\theta}_{l^{\prime}})+b_{nl^{\prime}}(\bar{\theta}_{l^{\prime}})\epsilon_{l^{\prime}})x_{l^{\prime}}\right]\notag\\
    &=\left(a_{nl}(\bar{\theta}_{l})+b_{nl}(\bar{\theta}_{l})\epsilon_{l}\right)x_{l\rightarrow n}+\underbrace{\sum\limits_{l^{\prime}=1,l^{\prime}\neq l}^L(a_{nl^{\prime}}(\bar{\theta}_{l^{\prime}})+b_{nl^{\prime}}(\bar{\theta}_{l^{\prime}})\epsilon_{l^{\prime}\rightarrow n})x_{l^{\prime}\rightarrow n}}_{{\mu_{\setminus l,n}}},&\label{mu}
\end{align}
\end{small}
\begin{small}
    \begin{align}
    &{\rm Var}\left[z_{n}|\epsilon_l\right]=\left(|a_{nl}(\bar{\theta}_{l})|^2+|b_{nl}(\bar{\theta}_{l})\epsilon_{l}|^2\right)\nu_{l\rightarrow n}^{x}+\underbrace{\sum\limits_{l^{\prime}=1,l^{\prime}\neq l}^L|b_{nl^{\prime}}(\bar{\theta}_{l^{\prime}})|^2\left(|x_{l^{\prime} \rightarrow n}|^2 \nu_{l^{\prime}\rightarrow n}^{\epsilon} \right.\left.+ |\epsilon_{l^{\prime}\rightarrow n}|^2\nu^x_{l^{\prime}\rightarrow n}\right)}_{v_{\setminus l,n}}\notag \\
    &\underbrace{+\sum\limits^L_{l^{\prime}=1,l^{\prime} \neq l}|b_{nl^{\prime}}(\bar{\theta}_{l^{\prime}})|^2 \nu^x_{l^{\prime}\rightarrow n}\nu^\epsilon_{l^{\prime}\rightarrow n} + |a_{nl^{\prime}}(\bar{\theta}_{l^{\prime}})|^2\nu^x_{l^{\prime}\rightarrow n}}_{v_{\setminus l,n}}.&\label{nu}
    \end{align}
\end{small}
Now we perform integration (\ref{origtheta}) with respect to $z_n$. Conditioned on $\epsilon_l$, by viewing
\begin{align}
	z_{n}|\epsilon_l\sim {\mathcal {CN}}(z_{n};{\rm E}\left[z_{n}|\epsilon_l\right],{\rm Var}\left[z_{n}|\epsilon_l\right]),\label{priorzm}
\end{align}
and
\begin{align}
    \tilde{y}_{n}=z_{n}+\tilde{w}_n,\label{llzm}
\end{align}
as the prior and likelihood of $z_n$, respectively, we perform integration over $z_{n}$ to yield
\begin{small}
    \begin{align}
    \tilde{y}_{n}\sim {\mathcal {CN}}(\tilde{y}_{n};\left(a_{nl}(\bar{\theta}_l)+b_{nl}(\bar{\theta}_l)\epsilon_l\right)x_{l\rightarrow n}+{\mu_{\setminus l,n}},{\rm Var}\left[z_{n}|\epsilon_l\right]+\tilde{\sigma}_n^2).
\end{align}
\end{small}
$f_{n}(\epsilon_l)$ (\ref{origtheta}) is calculated as
\begin{small}
    \begin{align}
		f_{n}(\epsilon_l)&=-\frac{|\tilde{y}_n-{\rm E}\left[z_n|\epsilon_l\right]|^2}{{\rm Var}\left[z_n|\epsilon_l\right] + \tilde{\sigma}^2_n} -\ln\left({\rm Var}\left[z_n|\epsilon_l\right]+\tilde{\sigma}^2_n\right) \notag\\
		& = \left(\epsilon_l^2-
		\frac{2\Re\left\{(\tilde{y}_n-\mu_{\setminus l,n})^*b_{nl}(\bar{\theta}_l)x_{l\rightarrow n}\epsilon_l\right\}}{|b_{nl}(\bar{\theta}_l)x_{l\rightarrow n}|^2}
		\right)/\left(\frac{{\rm Var}\left[z_n|\epsilon_l\right]+\tilde{\sigma}^2_n}{|b_{nl}(\bar{\theta}_l)x_{l\rightarrow n}|^2}\right)-\ln\left({\rm Var}\left[z_n|\epsilon_l\right]+\tilde{\sigma}^2_n\right) + {\rm const},\notag\\
		& = \left(\epsilon_l^2-
		2\Re\left\{\left(\frac{\tilde{y}_n-\mu_{\setminus l,n}}{b_{nl}(\bar{\theta}_l)x_{l\rightarrow n}}\right)^*\right\}\epsilon_l
		\right)/\left(\frac{{\rm Var}\left[z_n|\epsilon_l\right]+\tilde{\sigma}^2_n}{|b_{nl}(\bar{\theta}_l)x_{l\rightarrow n}|^2}\right)-\ln\left({\rm Var}\left[z_n|\epsilon_l\right]+\tilde{\sigma}^2_n\right)+{\rm const}.\label{fn_epsilon}
    \end{align}
\end{small}
Then we project $f_{n}(\epsilon_l)$ as a Gaussian distribution
\begin{align}\label{Projfn}
	{\rm Proj}\left[{\rm e}^{f_{n}(\epsilon_l)}\right]={\mathcal{N}}(\epsilon_l;\epsilon_{n\rightarrow l},\nu^\epsilon_{n\rightarrow l}),
\end{align}
where
\begin{align}
    \epsilon_{n\rightarrow l} = \Re\left\{ \frac{\tilde{y}_n-\mu_{\setminus l,n}}{b_{n,l}(\bar{\theta}_l) x_{l\rightarrow n}} \right\},\label{delta_mn}
\end{align}
$\nu^{\epsilon}_{n\rightarrow l}$ is calculated as
\begin{small}
    \begin{align}
    \nu^{\epsilon}_{n\rightarrow l} = & \left.\frac{{\rm Var}\left[z_n|\epsilon_l\right]+\tilde{\sigma}^2_n}{2|b_{n,l}(\bar{\theta}_l)x_{l\rightarrow n}|^2}\right|_{\epsilon_l = \epsilon_{n\rightarrow l}}\notag\\
    = & \frac{\tilde{\sigma}^2_n + \sum\limits_{l^{\prime}=1}^L|a_{nl^{\prime}}(\bar{\theta}_{l^{\prime}})|^2\nu^x_{l^{\prime}\rightarrow m} + \sum\limits_{l^{\prime}=1,l^{\prime}\neq l}^L |b_{nl^{\prime}}(\bar{\theta}_{l^{\prime}})|^2 \left(|x_{l^{\prime}\rightarrow n}|^2\nu^{\epsilon}_{l^{\prime}\rightarrow n} + |\epsilon_{l^{\prime}\rightarrow n}|^2 \nu^x_{l^{\prime}\rightarrow n} + \nu^{\epsilon}_{l^{\prime}\rightarrow n}\nu^x_{l^{\prime}\rightarrow n}\right)}{2|b_{n,l}(\bar{\theta}_l)x_{l\rightarrow n}|^2}.\label{sig_delta_mn}
    \end{align}
\end{small}
\begin{figure}[htbp]
	\centering
	\includegraphics[width=0.65\textwidth]{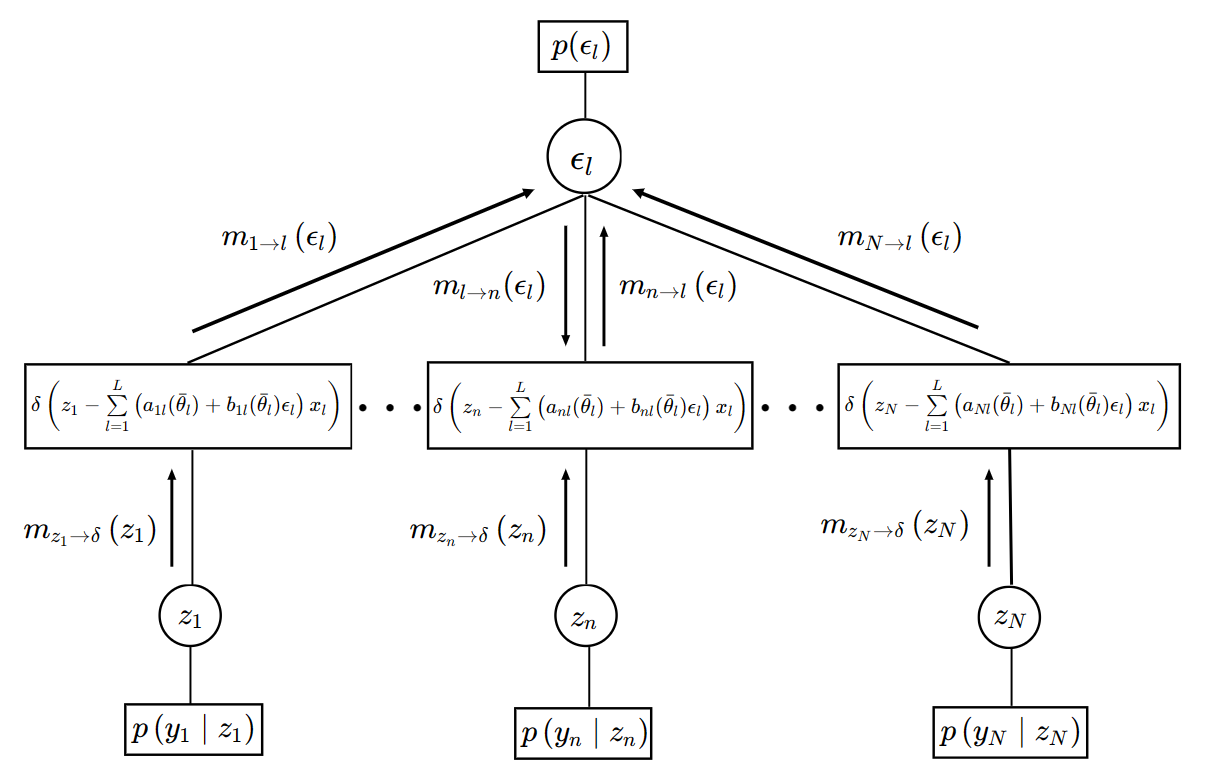}
	\caption{The factor graph for calculating ${m}_{l\rightarrow n}(\epsilon_l)$.}
	\label{equation25}
\end{figure}
\subsubsection{Updating of ${m}_{l\rightarrow n}(\epsilon_l)$}\label{mnthetan}
The subfactor graph is presented in Fig. \ref{equation25} and $m_{l\rightarrow n}(\epsilon_l)$ is updated as
\begin{align}\label{posttheta}
	m_{l\rightarrow n}(\epsilon_l)\propto \frac{{\rm Proj}\left[\prod\limits_{n=1}^N m_{n\rightarrow l}(\epsilon_l)p(\epsilon_l)\right]}{m_{n\rightarrow l}(\epsilon_l)}.
\end{align}
Given that the posterior mean and variance of variable $\epsilon_l$ is $\hat{\epsilon}_l$ and $\nu^{\epsilon}_l$ respectively, according to the definition in Table \ref{Notations},
\begin{subequations}
\begin{align}
	\frac{1}{\nu^{\epsilon}_{l\rightarrow n}} = \frac{1}{\nu^{\epsilon}_l} - \frac{1}{\nu^{\epsilon}_{n\rightarrow l}},\\
	\frac{\epsilon_{l\rightarrow n}}{\nu^{\epsilon}_{l\rightarrow n}} = \frac{\hat{\epsilon}_{l}}{\nu^{\epsilon}_l} - \frac{\epsilon_{n\rightarrow l}}{\nu^{\epsilon}_{n\rightarrow l}}\label{delta_nm}
\end{align}
\end{subequations}
are obtained. By the way, the pseudo measurement and variance of $\epsilon_l$ could be computed as 
\begin{subequations}
    \begin{align}
        \frac{1}{\nu^q_l} = \sum\limits^N_{n=1}\frac{1}{\nu^{\epsilon}_{n\rightarrow l}},\label{sig_delta_nm}\\
        \frac{q_l}{\nu^q_l} = \sum\limits^N_{n=1}\frac{\epsilon_{n\rightarrow l}}{\nu^{\epsilon}_{n\rightarrow l}}.
    \end{align}
\end{subequations}
\subsubsection{Updating of $m_{n\rightarrow l}(x_l)$}\label{tildemnxn}
\begin{figure}[htbp]
	\centering
	\includegraphics[width=0.65\textwidth]{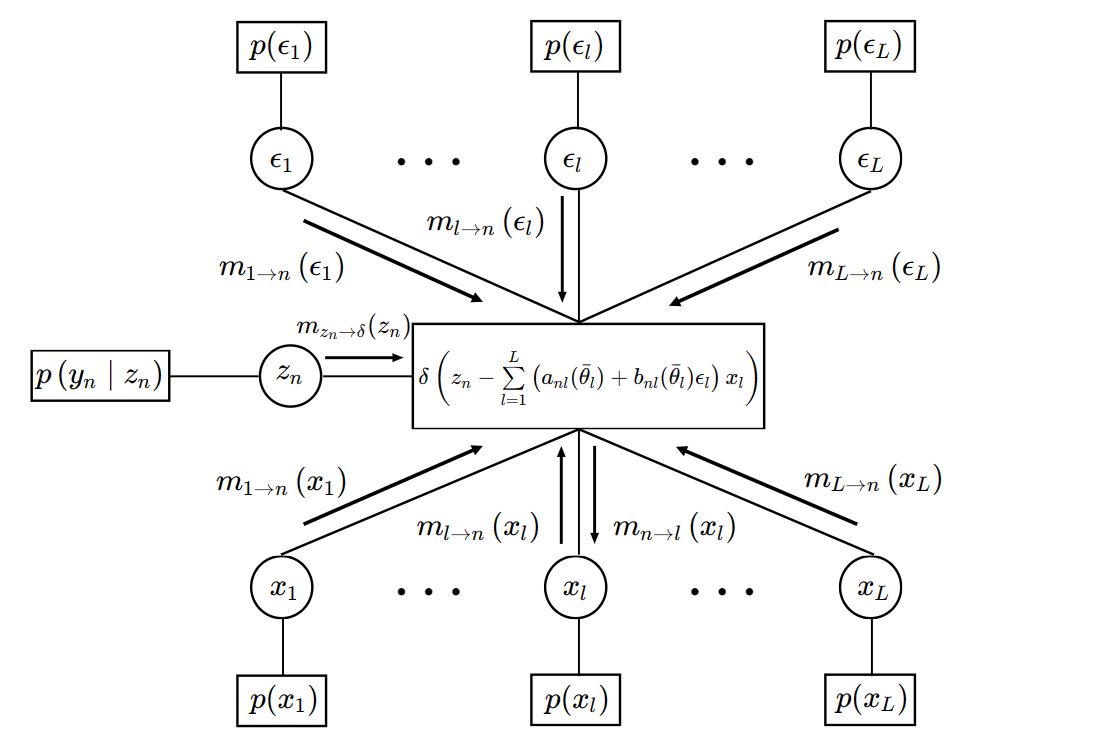}
	\caption{The factor graph for calculating $m_{n\rightarrow l}(x_l)$.}
	\label{equation27}
\end{figure}

The subfactor graph is presented in Fig. \ref{equation27}.
$m_{n\rightarrow l}(x_l)$ can be updated as
\begin{small}
    \begin{align}
        m_{n\rightarrow l}(x_l)&=\frac{{\rm {Proj}}\left[\int\prod\limits_{l=1}^L\left(m_{l\rightarrow n}(\epsilon_{l})m_{l\rightarrow n}(x_{l})\right)\delta\left(z_{n}-\left(\sum\limits_{l^{\prime}=1}^L(a_{nl^{\prime}}(\bar{\theta}_{l^{\prime}})+b_{nl^{\prime}}(\bar{\theta}_{l^{\prime}})\epsilon_{l^\prime})x_{l^{\prime}}\right)\right)m_{z_n\rightarrow\delta}(z_n)\prod\limits_{l=1}^L{\rm d}\epsilon_l\prod\limits_{l^{\prime}=1,l^{\prime}\neq l}^L{\rm d}x_{l^{\prime}}\right]}{{m}_{l\rightarrow n}(x_l)}\notag \\
		&=\frac{{\rm {Proj}}\left[m_{l\rightarrow n}(x_l){\rm e}^{g_n(x_l)}\right]}{m_{l\rightarrow n}(x_l)}.\label{origx1}
    \end{align}
\end{small}
Conditioned on $x_l$, we approximate $z_{n}$ as a Gaussian distribution by averaging over $\{x_{l^{\prime}}\}_{l^{\prime}\neq l}$ and $\{\epsilon_l\}_{l=1}^L$.
Straightforward calculation yields 
\begin{flalign}
    &{\rm E}\left[z_{n}|x_l\right]= \sum\limits^L_{l^{\prime}=1,l^{\prime}\neq l}{\rm E}\left[(a_{nl^{\prime}}(\bar{\theta}_{l^{\prime}})+b_{nl^{\prime}}(\bar{\theta}_{l^{\prime}})\epsilon_{l^{\prime}})x_{l^{\prime}}\right]+{\rm E}\left[a_{nl}(\bar{\theta}_l)+b_{nl}(\bar{\theta}_l)\epsilon_l
	\right]x_l \\
    &=\sum\limits^L_{l^{\prime}=1,l^{\prime}\neq l}\left(a_{nl^{\prime}}(\bar{\theta}_{l^{\prime}})+b_{nl^{\prime}}(\bar{\theta}_{l^{\prime}})\epsilon_{l^{\prime}\rightarrow n}\right)x_{l^{\prime}\rightarrow n}+ \underbrace{(a_{nl}(\bar{\theta}_l)+b_{nl}(\bar{\theta}_l)\epsilon_{l\rightarrow n})}_{\lambda_{l\rightarrow n}}x_l\\
    &=\lambda_{l\rightarrow n}x_l+{\mu_{\setminus l,n}},
\end{flalign}
\begin{align}
    &{\rm Var}\left[z_{n}|x_l\right]=\underbrace{|b_{nl}(\bar{\theta}_l)|^2\nu^{\epsilon}_{l\rightarrow n}}_{\gamma_{l\rightarrow n}}|x_l|^2 + \sum\limits^L_{l^{\prime}=1,l^{\prime}\neq l}|a_{nl}(\bar{\theta}_{l^{\prime}})|^2\nu^x_{l^{\prime}\rightarrow n} + \sum\limits^L_{l^{\prime}=1,l^{\prime} \neq l}|b_{nl^{\prime}}(\bar{\theta}_{l^{\prime}})|^2\left(|x_{l^{\prime}\rightarrow n}|^2\nu^{\epsilon}_{l^{\prime}\rightarrow n} \right.\notag\\ 
    & +\left.|\epsilon_{l^{\prime}\rightarrow n}|^2 \nu^x_{l^{\prime}\rightarrow n} + \nu^{\epsilon}_{l^{\prime}\rightarrow n}\nu^x_{l^{\prime}\rightarrow n}\right)\\
    &=\gamma_{l\rightarrow n}|x_l|^2+v_{\setminus l,n},
\end{align}
where $\mu_{\setminus l,n}$ and $\nu_{\setminus l,n}$ are defined in (\ref{mu}) and (\ref{nu}), respectively.

Thus, $g_n(x_l)$ (\ref{origx1}) is 
\begin{align}\label{gmxn}
    g_n(x_l)&=-\frac{|\tilde{y}_{n}-\mu_{\setminus l,n}-\lambda_{l\rightarrow n}x_l|^2}{{\rm Var}\left[z_{n}|x_l\right]+\tilde{\sigma}_n^2}-\ln\left({{\rm Var}\left[z_{n}|x_l\right]+\tilde{\sigma}_n^2}\right)+{\rm const}\notag\\
    &=-\frac{|\tilde{y}_{n}-\mu_{\setminus l,n}-\lambda_{l\rightarrow n}x_l|^2}{\gamma_{l\rightarrow n}|x_l|^2+v_{\setminus l,n}+\tilde{\sigma}_n^2}-\ln\left({\gamma_{l\rightarrow n}|x_l|^2+v_{\setminus l,n}+\tilde{\sigma}_n^2}\right)+{\rm const}.
\end{align}
Then we project ${m}_{l\rightarrow n}(x_l){\rm e}^{g_n(x_l)}$ as a Gaussian distribution. For simplicity, we first project ${\rm e}^{g_n(x_l)}$ as a Gaussian distribution. Then we have
\begin{align}
	m_{n\rightarrow l}(x_l)={\mathcal{CN}}(x_l;x_{n\rightarrow l},\nu^x_{n\rightarrow l})={\rm Proj}\left[{\rm e}^{g_n(x_l)}\right].
\end{align}
According to (\ref{gmxn}), we approximate $x_{n\rightarrow l}$ as
\begin{small}
    \begin{align}
    x_{n\rightarrow l}&=\frac{\tilde{y}_{n}-\mu_{\setminus l,n}}{\lambda_{l\rightarrow n}}\notag\\
    & = \frac{\tilde{y}_n-\sum\limits^L_{l^{\prime}=1,l^{\prime}\neq l}\left(a_{nl^{\prime}}(\bar{\theta}_{l^{\prime}})+b_{nl^{\prime}}(\bar{\theta}_{l^{\prime}})\epsilon_{l^{\prime}\rightarrow n}\right)x_{l^{\prime}\rightarrow n}}{a_{nl}(\bar{\theta}_l)+b_{nl}(\bar{\theta}_l)\epsilon_{l\rightarrow n}}, \label{x_mn}
\end{align}
\end{small}
and approximate $\nu^x_{n\rightarrow l}$ as (\ref{sigma_mn}),
\begin{figure*}[htbp]
    \begin{small}
    \begin{align}
    \nu^x_{n\rightarrow l}&=\frac{\gamma_{l\rightarrow n}|x_l|^2+v_{\setminus l,n}+\tilde{\sigma}_n^2}{|\lambda_{l\rightarrow n}|^2}\bigg|_{x_l=x_{n\rightarrow l}}\notag\\
    &= \frac{\sum\limits^L_{l^{\prime}=1,l^{\prime}\neq l}|b_{nl^{\prime}}(\bar{\theta}_{l^{\prime}})|^2\left(|x_{l^{\prime}\rightarrow n}|^2\nu^{\epsilon}_{l^{\prime}\rightarrow n} + |\epsilon_{l^{\prime}\rightarrow n}|^2 \nu^x_{l^{\prime}\rightarrow n} + \nu^{\epsilon}_{l^{\prime}\rightarrow n}\nu^x_{l^{\prime}\rightarrow n}\right) + \sum\limits^L_{l^{\prime}=1,l^{\prime}\neq l}|a_{nl^\prime}(\bar{\theta}_{l^\prime})|^2\nu^x_{l^{\prime}\rightarrow n} + |b_{nl}(\bar{\theta}_l)|^2\nu^\epsilon_{l\rightarrow n}|x_{n\rightarrow l}|^2 + \tilde{\sigma}^2_n}{|a_{nl}(\bar{\theta}_l)|^2 + |b_{nl}(\bar{\theta}_l)\epsilon_{l\rightarrow n}|^2} \label{sigma_mn}
    \end{align}
    \end{small}
\end{figure*}
where (\ref{sigma_mn}) utilizes $\Re\left\{a_{nl}(\bar{\theta}_l)b^*_{nl}(\bar{\theta}_l)\right\} = 0$.
\subsubsection{Updating of ${m}_{l\rightarrow n}(x_l)$}\label{mnxn}
\begin{figure}[t]
	\centering
	\includegraphics[width=0.65\textwidth]{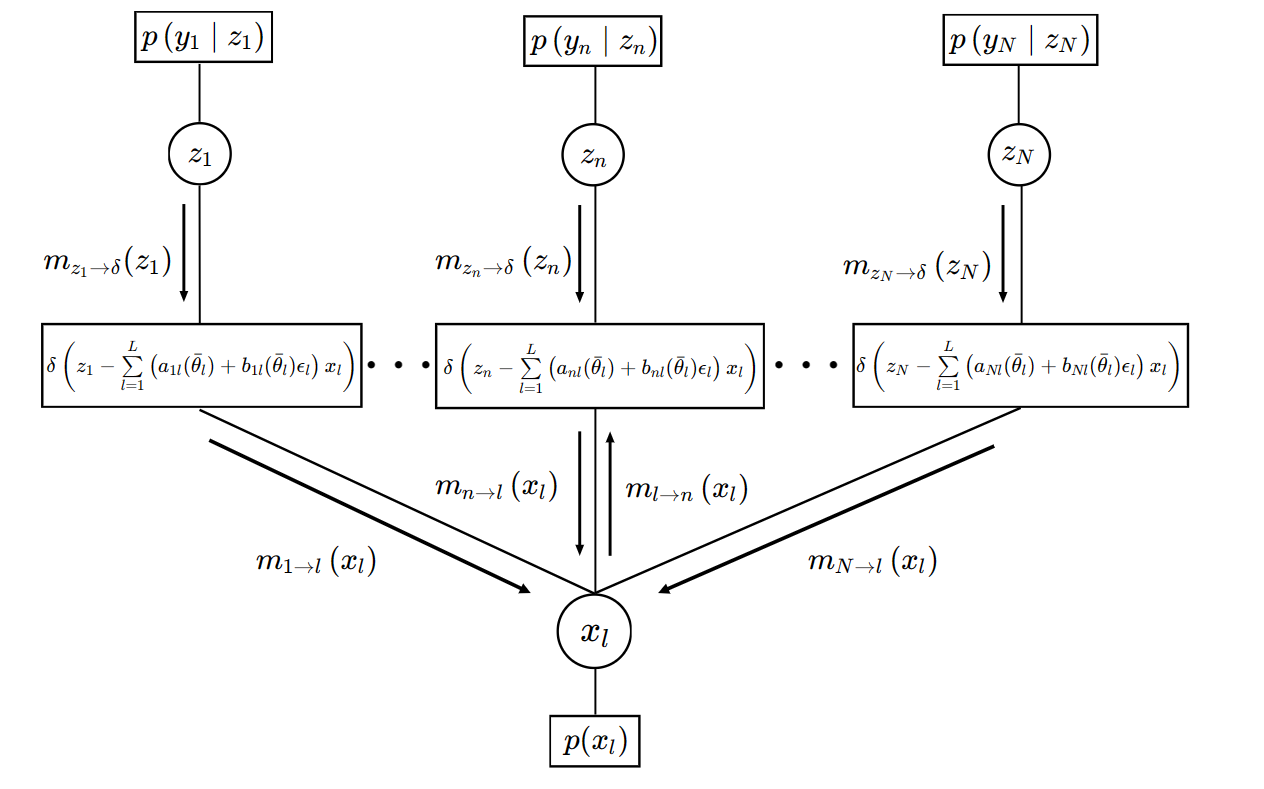}
	\caption{The factor graph for calculating ${m}_{l\rightarrow n}(x_l)$.}
	\label{equation8}
\end{figure}
The subfactor graph is presented in Fig. \ref{equation8}. According to EP, ${m}_{l\rightarrow n}(x_l)$ is updated as
\begin{align}
	m_{l\rightarrow n}(x_l)&\triangleq\mathcal{CN}(x_l;x_{l\rightarrow n},\nu^x_{l\rightarrow n}) \propto\frac{{\rm Proj}\left[\prod\limits_{n=1}^N m_{n\rightarrow l}(x_l)p(x_l)\right]}{m_{n\rightarrow l}(x_l)} \notag \\
	&\triangleq \frac{{\rm Proj}\left[{\mathcal {CN}}(x_l;r_l,\nu^r_l)p(x_l)\right]}{\mathcal{CN}(x_l;x_{n\rightarrow l},\nu^x_{n\rightarrow l})}.
\end{align}
where
\begin{subequations}
    \begin{align}
		\frac{1}{\nu^r_{l}}&=\sum\limits_{n=1}^N\frac{1}{\nu^x_{n\rightarrow l}},\label{sigma_n}\\
		\frac{r_l}{\nu^r_{l}}&=\sum\limits_{n=1}^N\frac{x_{n\rightarrow l}}{\nu^x_{n\rightarrow l}}\label{rlcompute}.
    \end{align}
\end{subequations}
Let
\begin{align}\label{hatmn}
	{\rm Proj}\left[{\mathcal {CN}}(x_l;r_l,\nu^r_n)p(x_l;{\boldsymbol \omega})\right]={\mathcal {CN}}(x_l;\hat{x}_l,\nu_l^x),
\end{align}
then
\begin{subequations}
    \begin{align}
		\frac{1}{\nu^x_{l\rightarrow n}}&=\frac{1}{\nu^x_l}-\frac{1}{\nu^x_{n\rightarrow l}},\label{sigmanm}\\
		\frac{x_{l\rightarrow n}}{\nu^x_{l\rightarrow n}}&=\frac{\hat{x}_l}{\nu^x_l}-\frac{x_{n\rightarrow l}}{\nu^x_{n\rightarrow l}}.\label{xnm}
	\end{align}
\end{subequations}
\subsubsection{Update of $m_{\delta\rightarrow z_n}(z_n)$}\label{mdeltaz}
\begin{figure}[htbp]
	\centering
	\includegraphics[width=0.65\textwidth]{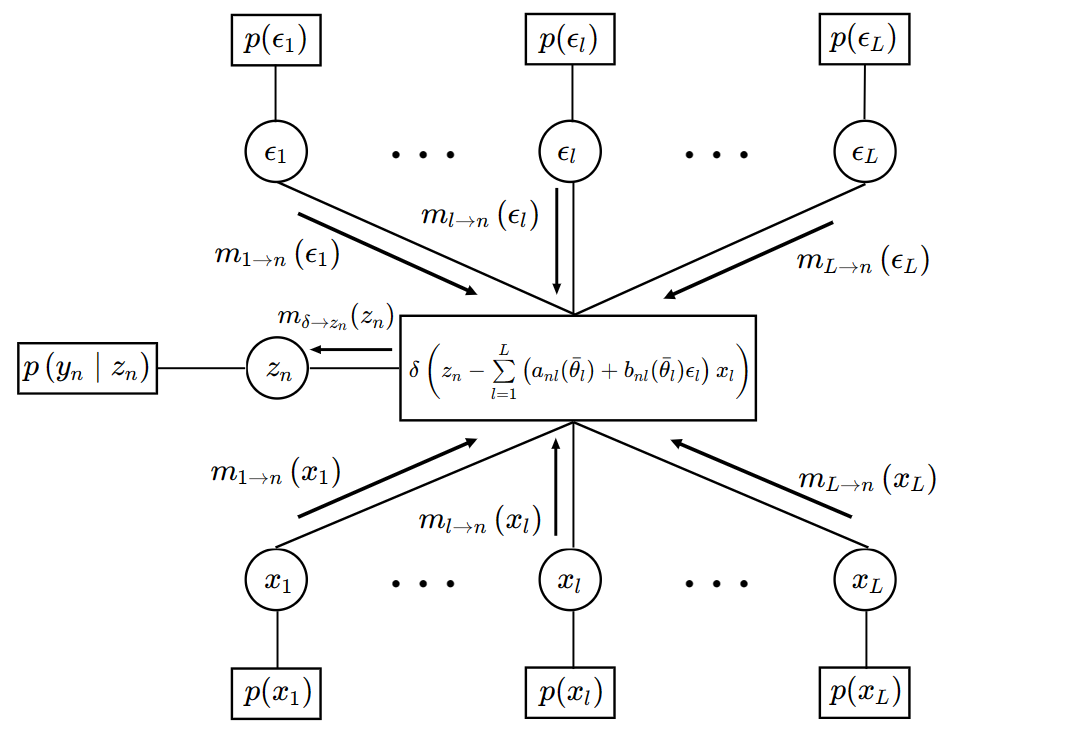}
	\caption{The factor graph for calculating $m_{\delta\rightarrow z_n}(z_n)$.}
	\label{deltazm}
\end{figure}
According to the factor graph in Fig. \ref{deltazm}, the message $m_{\delta\rightarrow z_n}(z_n)$ can be calculated as
\begin{small}
    \begin{align}\label{caldeltaz}
		m_{\delta\rightarrow z_n}(z_{n})=&\frac{{\rm Proj}\left[\int\prod\limits^L_{l=1}\left({m}_{l\rightarrow n}(\epsilon_l){m}_{l\rightarrow n}(x_l)\right)\delta\left(z_{n}-\sum\limits^L_{l^{\prime}=1}   
		(a_{nl^{\prime}}(\bar{\theta}_{l^{\prime}})+b_{nl^{\prime}}(\bar{\theta}_{l^{\prime}})\epsilon_{l^{\prime}}) x_{l^{\prime}}\right)m_{z_n\rightarrow \delta}(z_{n})\prod\limits_{l=1}^L{\rm d}x_l\prod\limits_{l=1}^L{\rm d}\epsilon_l\right]}{m_{z_n\rightarrow \delta}(z_{n})}\notag\\
		&=\frac{{\rm Proj}\left[q_{\rm A}(z_{n})m_{z_n\rightarrow \delta}(z_{n})\right]}{m_{z_n\rightarrow \delta}(z_{n})}
    \end{align}
\end{small}
We first approximate $q_{\rm A}(z_{n})$ as a Gaussian distribution, then we have
\begin{align}\label{caldeltaz1}
m_{\delta\rightarrow z_n}(z_{n})={\rm Proj}\left[q_{\rm A}(z_{n})\right].
\end{align}
The means and variances of $z_{n}$ with respect to $q_{\rm A}(z_{n})$ can be calculated as
\begin{align}
    {z}_{{\rm A},n}^{\rm ext}(t)&=\sum\limits_{l=1}^L{\rm E}\left[(a_{nl}(\bar{\theta}_l)+b_{nl}(\bar{\theta}_l)\epsilon_l)x_l\right]=\sum\limits_{l=1}^L(a_{nl}(\bar{\theta}_l)+b_{nl}(\bar{\theta}_l)\epsilon_{l\rightarrow n})x_{l\rightarrow n},\\
    {v}_{{\rm A},n}^{\rm ext}(t)&= \sum\limits_{l=1}^L |b_{nl}(\bar{\theta}_l)|^2 \left(|x_{l\rightarrow n}|^2\nu^{\epsilon}_{l\rightarrow n} + |\epsilon_{l\rightarrow n}|^2 \nu^x_{l\rightarrow n}\right) + \sum\limits_{l=1}^L|b_{nl}(\bar{\theta}_l)|^2\nu^{\epsilon}_{l\rightarrow n}\nu^x_{l\rightarrow n} +\sum\limits_{l=1}^L|a_{nl}(\bar{\theta}_l)|^2\nu^x_{l\rightarrow n}.
\end{align}
Using (\ref{caldeltaz1}) we obtain $m_{\delta\rightarrow z_n}(z_n)$ as
\begin{align}
	m_{\delta\rightarrow z_n}(z_n)={\mathcal {CN}}({z}_n;z_{{\rm A},n}^{\rm ext},{v}_{{\rm A},n}^{\rm ext}).
\end{align}
Now the message passing algorithm is closed.
\subsection{Simplifying the Message as BiG-LSE}
The maximum number of the above messages transmitted is $LN$, which is too large in the high dimensional setting. As a consequence, some approximations are adopted to obtain the simplified BiG-LSE, whose maximum number of transmitted messages is $L+N$.

Now we define 
\begin{align}
	\hat{s}_n \triangleq \frac{\tilde{y}_n-\hat{p}_n}{\nu^p_n+\tilde{\sigma}^2_n},\label{shat}\\
	\nu^s_n\triangleq\frac{1}{\nu^p_n+\tilde{\sigma}^2_n}\label{tau_s},
\end{align}
where $\hat{p}_n$ and $\nu^p_n$ are
\begin{align}
    \hat{p}_n & \triangleq {z}_{{\rm A},n}^{\rm ext} = \sum\limits_{l=1}^L a_{l\rightarrow n}x_{l\rightarrow n}, \label{phat}\\
    \nu^p_n & \triangleq {v}_{{\rm A},n}^{\rm ext} = \sum\limits_{l=1}^L|a_{nl}(\bar{\theta}_l)|^2\nu^x_{l\rightarrow n} + |b_{nl}(\bar{\theta}_l)|^2|x_{l\rightarrow n}|^2\nu^{\epsilon}_{l\rightarrow n}+ \sum\limits_{l=1}^L|b_{nl}(\bar{\theta}_l)|^2\left(|\epsilon_{l\rightarrow n}|^2 \nu^{x}_{l\rightarrow n} + \nu^{\epsilon}_{l\rightarrow n}\nu^x_{l\rightarrow n}\right),\label{tau_m_p}
\end{align}
where $a_{l\rightarrow n}$ is defined as
\begin{equation}
	a_{l\rightarrow n} \triangleq a_{nl}(\bar{\theta}_{l})+b_{nl}(\bar{\theta}_{l})\epsilon_{l\rightarrow n}. \label{a_ln}
\end{equation}
\subsubsection{Approximate $a_{l\rightarrow n}$ and $x_{l\rightarrow n}$}
During iteration, $\nu^\epsilon_l$ approximates $\nu^\epsilon_{l\rightarrow n}$ well, i.e., 
\begin{equation}
	\nu^\epsilon_{l\rightarrow n} \approx \nu^\epsilon_l.\label{tau_delta_ln}
\end{equation}
Substituting (\ref{tau_delta_ln}) into (\ref{delta_nm}), $\epsilon_{l\rightarrow n}$ can be approximated as
\begin{equation}
	\epsilon_{l\rightarrow n} = \hat{\epsilon}_l - \nu^\epsilon_l\frac{\epsilon_{n\rightarrow l}}{\nu^{\epsilon}_{n\rightarrow l}}\label{delta_nm2}
\end{equation}
Substituting (\ref{delta_mn}) and (\ref{sig_delta_mn}) into $\frac{\epsilon_{n\rightarrow l}}{\nu^{\epsilon}_{n\rightarrow l}}$, we have
\begin{align}
    \frac{\epsilon_{n\rightarrow l}}{\nu^\epsilon_{n\rightarrow l}}&=\frac{2\Re\left\{\left(\tilde{y}_n-\sum\limits^L_{l^{\prime}\neq l}\left[a_{nl^{\prime}}(\bar{\theta}_{l^{\prime}})+b_{nl^{\prime}}(\bar{\theta}_{l^{\prime}})\epsilon_{l^{\prime}\rightarrow n}\right]x_{l^{\prime}\rightarrow n}\right)b^{*}_{nl}(\bar{\theta}_l)x^*_{l\rightarrow n}\right\}}{\tilde{\sigma}^2_n + \sum\limits_{l^{\prime}=1}^L|a_{nl^{\prime}}(\bar{\theta}_{l^{\prime}})|^2\nu^x_{l^{\prime}\rightarrow m} + \sum\limits_{l^{\prime}=1,l^{\prime}\neq l}^L |b_{nl^{\prime}}(\bar{\theta}_{l^{\prime}})|^2 \left(|x_{l^{\prime}\rightarrow n}|^2\nu^{\epsilon}_{l^{\prime}\rightarrow n} + |\epsilon_{l^{\prime}\rightarrow n}|^2 \nu^x_{l^{\prime}\rightarrow n} + \nu^{\epsilon}_{l^{\prime}\rightarrow n}\nu^x_{l^{\prime}\rightarrow n}\right)}\notag\\
    &\approx 2\Re\left\{\hat{s}_n b^{*}_{nl}(\bar{\theta}_l)x^*_{l\rightarrow n}\right\} \label{deltaovermuntol}
\end{align}
Substituting (\ref{delta_nm2}) and (\ref{deltaovermuntol}) into (\ref{a_ln}), $a_{l\rightarrow n}$ can be approximated as
\begin{align}
    a_{l\rightarrow n} & \approx a_{nl}(\bar{\theta}_l) + b_{nl}(\bar{\theta}_l)\hat{\epsilon}_l - 2b_{nl}(\bar{\theta}_l)\nu^\epsilon_l\Re\left\{{b^*_{nl}(\bar{\theta}_l)\hat{s}_n x^*_{l\rightarrow n}}\right\} \notag\\
    & = \hat{a}_{nl} - 2b_{nl}(\bar{\theta}_l)\nu^\epsilon_l\Re\left\{{b_{nl}(\bar{\theta}_l)\hat{s}^*_n x_{l\rightarrow n}}\right\},\label{a_nm}
\end{align}
where $\hat{a}_{nl} = a_{nl}(\bar{\theta}_l) + b_{nl}(\bar{\theta}_l)\hat{\epsilon}_l$.

Similarly, during iteration, $\nu^x_l$ approximates $\nu^x_{l\rightarrow n}$ well, i.e., 
\begin{equation}
	\nu^x_{l\rightarrow n} \approx \nu^x_l.\label{tau_x_ln}
\end{equation}
Substituing (\ref{tau_x_ln}) into (\ref{xnm}), $x_{l\rightarrow n}$ can be approximated as
\begin{align}
	x_{l\rightarrow n} &\approx \hat{x}_l-\nu^x_l\frac{x_{n\rightarrow l}}{\nu^x_{n\rightarrow l}}.\label{x_nm}
\end{align}
Substituting $x_{n\rightarrow l}$ (\ref{x_mn}) and $\nu^x_{n\rightarrow l}$ (\ref{sigma_mn}) into $\frac{x_{n\rightarrow l}}{\nu^x_{n\rightarrow l}}$, one has
\begin{align}
    \frac{x_{n\rightarrow l}}{\nu^x_{n\rightarrow l}}&= \frac{\left(\tilde{y}_n-\sum\limits^L_{l^{\prime}=1,l^{\prime}\neq l}\left(a_{nl^{\prime}}(\bar{\theta}_{l^{\prime}})+b_{nl^{\prime}}(\bar{\theta}_{l^{\prime}})\epsilon_{l^{\prime}\rightarrow n}\right)x_{l^{\prime}\rightarrow n}\right)(a_{nl}(\bar{\theta}_l) + b_{nl}(\bar{\theta}_l)\epsilon_{l\rightarrow n})^\star}{\sum\limits^L_{l^{\prime}=1,l^{\prime}\neq l}|b_{nl^{\prime}}(\bar{\theta}_{l^{\prime}})|^2\left(|x_{l^{\prime}\rightarrow n}|^2\nu^{\epsilon}_{l^{\prime}\rightarrow n} + |\epsilon_{l^{\prime}\rightarrow n}|^2 \nu^x_{l^{\prime}\rightarrow n} + \nu^{\epsilon}_{l^{\prime}\rightarrow n}\nu^x_{l^{\prime}\rightarrow n}\right) + \sum\limits^L_{l^{\prime}=1,l^{\prime}\neq l}|a_{nl^\prime}(\bar{\theta}_{l^\prime})|^2\nu^x_{l^{\prime}\rightarrow n} + \tilde{\sigma}^2_n} \notag\\
    &=\frac{\tilde{y}_n-\hat{p}_n}{\tilde{\sigma}^2_n+\nu^p_n}a^*_{l\rightarrow n}=\hat{x}_l-\nu^x_l\hat{s}_n a^*_{l\rightarrow n}\label{xnl_over_taunl} 
\end{align}	
Substituting (\ref{xnl_over_taunl}) into (\ref{x_nm}), $x_{l\rightarrow n}$ can be approximated as
\begin{align}
	x_{l\rightarrow n}\approx \hat{x}_l-\nu^x_l\frac{\tilde{y}_n-\hat{p}_n}{\tilde{\sigma}^2_n+\nu^p_n}a^*_{l\rightarrow n}=\hat{x}_l-\nu^x_l\hat{s}_n a^*_{l\rightarrow n}.\label{x_nmapp}
\end{align}	
\subsubsection{Approximate $\hat{p}_n$ and $\nu^p_n$}
Substituting $a_{l\rightarrow n}$ (\ref{a_nm}) and $x_{l\rightarrow n}$ (\ref{x_nmapp}) into (\ref{phat}), $\hat{p}_n$ can be approximated as 
\begin{small}
   \begin{align}
    & \hat{p}_n = \sum\limits_l a_{l\rightarrow n}x_{l\rightarrow n}\notag\\
    & \approx \sum\limits_l^L\left(\hat{a}_{nl} - 2b_{nl}(\bar{\theta}_l)\nu^\epsilon_l\Re\left\{{b_{nl}(\bar{\theta}_l)\hat{s}^*_n x_{l\rightarrow n}}\right\}\right)\left(\hat{x}_l-\nu^x_l\hat{s}_na^*_{l\rightarrow n}\right)\notag\\
    & \stackrel{a}\approx \sum\limits_l^L\left(\hat{a}_{nl} - 2b_{nl}(\bar{\theta}_l)\nu^\epsilon_l\Re\left\{{b_{nl}(\bar{\theta}_l)\hat{s}^*_n \hat{x}_{l}}\right\}\right)\left(\hat{x}_l-\nu^x_l\hat{s}_n\hat{a}^*_{nl}\right)\notag\\
    & \approx \sum\limits_l^L\hat{a}_{nl}\hat{x}_l-\hat{s}_n\sum\limits_l^L|\hat{a}_{nl}|^2 \nu^x_l - 2\sum\limits^L_l \nu^{\epsilon}_{l}b_{nl}(\bar{\theta}_l)\Re\left\{b^*_{nl}(\bar{\theta}_l)\hat{s}_n\hat{x}^*_l\right\}\hat{x}_l\notag\\
    & \approx \sum\limits_l^L\hat{a}_{nl}\hat{x}_l-\hat{s}_n\sum\limits_l^L|\hat{a}_{nl}|^2 \nu^x_l - \sum\limits^L_l \nu^{\epsilon}_{l}|b_{nl}(\bar{\theta}_l)|^2\hat{s}_n|\hat{x}_l|^2,\label{phatn4}
\end{align} 
\end{small}
where $\stackrel{a}\approx$ is due to $ x_{l\rightarrow n}\approx  \hat{x}_{l}$ and $a_{l\rightarrow n}\approx \hat{a}_{nl}$.

Furthermore, by defining $\nu^a_{nl}$ as 
\begin{align}
\nu^a_{nl} = |b_{nl}(\bar{\theta}_l)|^2 \nu^{\epsilon}_l,
\end{align}
$\hat{p}_n$ (\ref{phatn4}) can be rewritten as
\begin{align}
    \hat{p}_n \approx \sum\limits_l^L\hat{a}_{nl}\hat{x}_l-\hat{s}_n\sum\limits_l^L|\hat{a}_{nl}|^2 \nu^x_l - \hat{s}_n\sum\limits^L_l \nu^a_{nl}|\hat{x}_l|^2,
\end{align}
$\nu^p_n$ (\ref{tau_m_p}) can be approximated as 
\begin{align}
\nu^p_n &= \sum\limits_{l=1}^L |b_{nl}(\bar{\theta}_l)|^2\left(|x_{l\rightarrow n}|^2 \nu^{\epsilon}_{l\rightarrow n} + |\epsilon_{l\rightarrow n}|^2 \nu^x_{l\rightarrow n} + \nu^{\epsilon}_{l\rightarrow n}\nu^x_{l\rightarrow n}\right)+ \sum\limits_{l=1}^L|{a}_{nl}(\bar{\theta}_l)|^2\nu^x_{l\rightarrow n} ,\\
& \stackrel{b}\approx \sum\limits^L_{l=1}|\hat{a}_{nl}|^2 \nu^x_l + \sum\limits^L_{l=1}|\hat{x}_l|^2 \nu^a_{nl} + \sum\limits^L_{l=1} \nu^x_l\nu^a_{nl}.
\end{align}
where $\stackrel{b}\approx$ is due to $\nu^x_{l\rightarrow n}\approx \nu^x_l$, $\nu^{\epsilon}_{l\rightarrow n}\approx \nu^\epsilon_l$, $|x_{l\rightarrow n}|^2 \approx |\hat{x}_l|^2$, $|\epsilon_{l\rightarrow n}|^2\approx |\hat{\epsilon}_l|^2$, and $|\hat{a}_{nl}|^2=|{a}_{nl}(\bar{\theta}_l)|^2+|b_{nl}(\bar{\theta}_l)|^2\hat{\epsilon}_l^2$.

\subsubsection{Approximate $\hat{r}_l$, $\nu^r_l$, $\hat{q}_l$ and $\nu^q_l$}
According to (\ref{sigma_n}) and (\ref{sigma_mn}), $\nu^r_l$ can be computed as
\begin{align}
    \frac{1}{\nu^r_l} &= \sum\limits^N_{n=1}\frac{1}{\nu^x_{n\rightarrow l}}  = \sum\limits^N_{n=1}\frac{|\lambda_{l\rightarrow n}|^2}{\gamma_{l\rightarrow n}|x_{n\rightarrow l}|^2+\nu_{\setminus l,n}+\tilde{\sigma}^2_n},\\
    & \stackrel{c}\approx \sum\limits^N_{n=1}\frac{|a_{nl}(\bar{\theta}_l)|^2+|b_{nl}(\bar{\theta}_l)\hat{\epsilon}_l|^2}{\tilde{\sigma}^2_n+\nu^p_n} \approx \sum\limits^N_{n=1}|\hat{a}_{nl}|^2\nu^s_n.
\end{align}
where $\stackrel{c}\approx$ is due to $\epsilon_{l\rightarrow n}\approx\hat{\epsilon}_l$ and $|\lambda_{l\rightarrow n}|^2=|{a}_{nl}(\bar{\theta}_l)+b_{nl}(\bar{\theta}_l)\epsilon_{l\rightarrow n}|^2\approx |\hat{a}_{nl}|^2=|{a}_{nl}(\bar{\theta}_l)|^2+|b_{nl}(\bar{\theta}_l)|^2\hat{\epsilon}_l^2$.

Computing ${r}_l$ (\ref{rlcompute}) involves calculating $\frac{x_{n\rightarrow l}}{\nu^x_{n\rightarrow l}}$. Substituting $x_{n\rightarrow l}$ (\ref{x_mn}) and $\nu^x_{n\rightarrow l}$ (\ref{sigma_mn}) into $\frac{x_{n\rightarrow l}}{\nu^x_{n\rightarrow l}}$ yields 
\begin{align}
    \frac{x_{n\rightarrow l}}{\nu^x_{n\rightarrow l}} &= a^*_{l\rightarrow n}\frac{\tilde{y}_n-\hat{p}_n+a_{l\rightarrow n}x_{l\rightarrow n}}{\tilde{\sigma}^2_n + \nu^p_n}\notag\\
    & =a^*_{l\rightarrow n}\hat{s}_n+|a_{l\rightarrow n}|^2x_{l\rightarrow n}\nu^s_n\\
    &\stackrel{f}\approx\left(\hat{a}_{nl} - 2b_{nl}(\bar{\theta}_l)\nu^{\epsilon}_l{\rm Re}\left\{b^*_{nl}(\bar{\theta}_l)\hat{s}_n\hat{x}^*_l\right\}\right)^*\hat{s}_n + |\hat{a}_{nl}|^2\hat{x}_l \nu^s_n\notag\\
    &\approx \hat{a}_{nl}^*\hat{s}_n-|b_{nl}(\bar{\theta}_l)|^2|\hat{s}_n|^2\nu^{\epsilon}_l\hat{x}_l+ |\hat{a}_{nl}|^2\hat{x}_l \nu^s_n.
\end{align}
where $a_{l\rightarrow n}$ has been defined in (\ref{a_ln}), $\stackrel{f}\approx$ is due to (\ref{a_nm}), $x_{l\rightarrow n}\approx \hat{x}_l$ and $|a_{l\rightarrow n}|^2\approx |\hat{a}_{nl}|^2$. Consequently, ${r}_l$ can be approximated as 
\begin{align}
    {r}_l &= \nu^r_l\sum\limits^N_{n=1}\frac{x_{n\rightarrow l}}{\nu^x_{n\rightarrow l}}\\
    &\approx \nu^r_l\sum\limits^N_{n=1}\hat{a}^*_{nl}\hat{s}_n - 2\nu^r_l\nu^\epsilon_l\sum\limits^N_{n=1} b^*_{nl}(\bar{\theta}_l)\Re\left\{b_{nl}(\bar{\theta}_l)\hat{x}_l\hat{s}^*_n\right\}\hat{s}_n+ \hat{x}_l\nu^r_l\sum\limits^N_{n=1}|\hat{a}_{nl}|^2\nu^s_n\\
    &=\hat{x}_l + \nu^r_l\sum\limits^N_{n=1}\hat{a}^*_{nl}\hat{s}_n -2\nu^r_l\nu^\epsilon_l\sum\limits^N_{n=1} b^*_{nl}(\bar{\theta}_l)\Re\left\{b_{nl}(\bar{\theta}_l)\hat{x}_l\hat{s}^*_n\right\}\hat{s}_n\notag\\
    &\approx \hat{x}_l + \nu^r_l\sum\limits^N_{n=1}\hat{a}^*_{nl}\hat{s}_n - \nu^r_l\hat{x}_l\sum^N_{n=1}|\hat{s}_n|^2 \nu^a_{nl}.
\end{align}
Now we compute $\hat{q}_l$ and $\nu^q_l$. According to (\ref{sig_delta_nm}) and definition of $\nu^p_n$ (\ref{tau_m_p}), $\nu^q_l$ can be caculated as
\begin{align}\label{nuql}
    \frac{1}{\nu^q_l} & = \sum\limits^N_{n=1}\frac{1}{\nu^{\epsilon}_{n\rightarrow l}}\approx \sum\limits^N_{n=1}\frac{2|b_{nl}(\bar{\theta}_l)x_{l\rightarrow n}|^2}{\tilde{\sigma}^2_n+\nu^p_n}\\
    & \stackrel{g}\approx 2\sum\limits^N_{n=1}|b_{nl}(\bar{\theta}_l)|^2|\hat{x}_l|^2 \nu^s_n.
\end{align}
where $\stackrel{g}\approx$ is due to $|x_{l\rightarrow n}|^2 \approx |\hat{x}_l|^2$ and the definition of $\nu^s_n$ (\ref{tau_s}).
\begin{align}
    \hat{q}_l & = \nu^q_l\sum\limits^N_{n=1}\frac{\epsilon_{n\rightarrow l}}{\nu^{\epsilon}_{n\rightarrow l}} \notag\\
    & \approx \nu^q_l\sum\limits^N_{n=1}\frac{2\Re\left\{\left(\tilde{y}_n-\hat{p}_n+a_{l\rightarrow n}x_{l\rightarrow n}\right) b^*_{nl}(\bar{\theta}_l)x^*_{l\rightarrow n}\right\}}{\nu^p_n+\tilde{\sigma}^2_n}\\
    & = \nu^q_l\sum\limits^N_{n=1} 2\Re\left\{\hat{s}_n\left(b_{nl}(\bar{\theta}_l)x_{l\rightarrow n}\right)^*\right\} \nu^q_l\sum\limits^N_{n=1}\frac{2|x_{l\rightarrow n}|^2\Re\left\{a_{l\rightarrow n} b^*_{nl}(\bar{\theta}_l)\right\}}{\nu^p_n+\tilde{\sigma}^2_n}
\end{align}
\begin{align}
    & \stackrel{h}\approx \nu^q_l\sum\limits^N_{n=1} 2\Re\left\{\hat{s}_n\left(b_{nl}(\bar{\theta}_l)\hat{x}_l\right)^*\right\}+ \nu^q_l\sum^N_{n=1} \nu^s_n\left(2|b_{nl}(\bar{\theta}_l)|^2\hat{\epsilon}_l + 2\Re\left\{b^{*}_{nl}(\bar{\theta}_l)a_{nl}(\bar{\theta}_l)\right\}\right)|\hat{x}_l|^2\notag\\
    & \stackrel{i}\approx \hat{\epsilon}_l + \nu^q_l\sum\limits^N_{n=1}2\Re\left\{\hat{s}_n\left(b_{nl}(\bar{\theta}_l)\hat{x}_l\right)^*\right\},
\end{align}
where $\stackrel{h}\approx$ is due to $a_{l\rightarrow n} \approx \hat{a}_{nl}=a_{nl}(\bar{\theta}_{l})+b_{nl}(\bar{\theta}_{l})\hat{\epsilon}_l$ and $x_{l\rightarrow n}\approx \hat{x}_l$, $\stackrel{i}=$ is due to (\ref{nuql}) and $\Re\left\{b^{*}_{nl}(\bar{\theta}_l)a_{nl}(\bar{\theta}_l)\right\}=0$.
We summarize the scalar form of BiG-LSE algorithm shown in Algorithm \ref{BiGLSEscalar}.
\begin{algorithm}[htbp]
	\caption{BiGLSE (scalar form)}
	\begin{algorithmic}[1]\label{BiGLSEscalar}
		\STATE Definition:\\
		\STATE $p_{z_n \mid p_n}\left(z \mid \hat{p} ; \nu^p\right)  \triangleq \frac{p_{y_n \mid z_n}\left(y \mid z\right) \mathcal{CN}\left(z;\hat{p}, \nu^p\right)}{\int_{z} p_{y_n \mid z_n}\left(y \mid z\right) \mathcal{CN}\left(z; \hat{p}, \nu^p\right){\rm d}z}$ \\
		\STATE $p_{x_l \mid r_l}\left(x \mid \hat{r} ; \nu^r\right) \triangleq \frac{p_{x_l}(x) \mathcal{CN}\left(x ; \hat{r}, \nu^r\right)}{\int_{x} p_{x_l}\left(x\right) \mathcal{CN}\left(x ; \hat{r}, \nu^r\right){\rm d}x}$ \\
		\STATE $p_{\epsilon_l\mid q_l}\left(\epsilon \mid \hat{q} ; \nu^q\right) \triangleq \frac{p_{\mathrm{\epsilon}_l}(\epsilon) \mathcal{N}\left(\epsilon ; \hat{q}, \nu^q\right)}{\int_{\epsilon} p_{\mathrm{\epsilon}_l}\left(\epsilon\right) \mathcal{N}\left(\epsilon ; \hat{q}, \nu^q\right){\rm d}\epsilon}$\\
		\STATE Initialization:\\
		\STATE $\forall n: \hat{s}_{n}(0)=0 $ \\
		\STATE $\forall l: \rm{set}\ \hat{x}_l(1),\nu^x_l(1),\hat{\epsilon}_l(1),\nu^\epsilon_l(1)$
		\FOR{$t=1, \ldots T_{\max }$}
		\STATE $\forall n,l: \hat{a}_{nl}(t) =  a_{nl}(\bar{\theta}_l) + b_{nl}(\bar{\theta}_l)\hat{\epsilon}_l(t)$\\
		\STATE $\forall n,l: \nu^a_{nl}(t) = |b_{nl}(\bar{\theta}_l)|^2 \nu^\epsilon_l(t)$ \\
		\STATE $\forall n: \bar{\nu}^p_n(t) = \sum_{l}|\hat{a}_{nl}(t)|^2 \nu^x_l(t) + \sum_{l} \nu^a_{nl}(t)|\hat{x}_l(t)|^2$\\
		\STATE $\forall n: \nu^p_n(t) = \bar{\nu}^p_n(t) + \sum_{l} \nu^x_l(t) \nu^a_{nl}(t) $\\
		\STATE $\forall n: \hat{p}_n(t) = \sum_{l}\hat{a}_{nl}(t)\hat{x}_l(t)-\hat{s}_n(t)\sum_{l}|\hat{a}_{nl}(t)|^2 \nu^x_l(t) - \hat{s}_n(t)\sum_{l} \nu^a_{nl}(t)|\hat{x}_l(t)|^2$ \\
		\STATE $\forall n: \nu^z_n(t)={\mathrm {Var}}\left\{{z}_{n} \mid {p}_{n}=\hat{p}_{n}(t) ; \nu_{n}^p(t)\right\}$\\
		\STATE $\forall n: \hat{z}_{n}(t)={\mathrm E}\left\{{z}_{n} \mid {p}_{n}=\hat{p}_{n}(t) ; \nu_{n}^p(t)\right\}$\\
		\STATE $\forall n: \nu_{n}^s(t)=\left(1-\nu_{n}^z(t) / \nu_{n}^p(t)\right) / \nu_{n}^p(t)$\\
		\STATE $\forall n: \hat{s}_{n}(t)=\left(\hat{z}_{n}(t)-\hat{p}_{n}(t)\right) / \nu_{n}^p(t)$\\
		\STATE $\forall l: \nu^r_l(t)=\left(\sum_{n}|\hat{a}_{nl}(t)|^2\nu_{n}^s(t) \right)^{-1}$\\
		\STATE $\forall l: \hat{r}_l(t)=\hat{x}_l(t) + \nu^r_l(t)\sum_{n}\hat{a}^*_{nl}(t)\hat{s}_n(t) - \nu^r_l(t)\hat{x}_l(t)\sum_{n}\nu^a_{nl}(t)|\hat{s}_n(t)|^2 $\\
		\STATE $\forall l: \nu_l^q(t) = \left(2\sum_{n}|b_{nl}(\bar{\theta}_l)\hat{x}_l(t)|^2 \nu_{n}^s(t)\right)^{-1}$\\
		\STATE $\forall l: \hat{q}_l(t)=\hat{\epsilon}_l(t) + \nu_l^q(t)\sum_{n}2{\rm Re}\left\{\hat{s}_n(t)b^*_{nl}(\bar{\theta}_l)\hat{x}^*_l(t)\right\}$\\
		\STATE $\forall l: \nu_l^x(t+1)=\mathrm{Var}\left\{x_l \mid r_l=\hat{r}_l(t) ; \nu_l^r(t)\right\}$\\
		\STATE $\forall l: \hat{x}_l(t+1)=\mathrm{E}\left\{x_l \mid r_l=\hat{r}_l(t) ; \nu_l^r(t)\right\}$\\
		\STATE $\forall l: \nu^\epsilon_l(t+1)=\mathrm{Var}\left\{\epsilon_l \mid q_l=\hat{q}_l(t) ; \nu_l^q(t)\right\}$\\
		\STATE $\forall l: \hat{\epsilon}_l(t+1)=\mathrm{E}\left\{\epsilon_l \mid q_l=\hat{q}_l(t) ; \nu_l^q(t)\right\}$\\
		\ENDFOR
    \end{algorithmic}
\end{algorithm}

The matrix-vector version: Let $\bar{{\mathbf A}}(t)$ and $\bar{{\mathbf B}}(t)$ denote the $t$th iteration of ${\mathbf A}(\bar{\boldsymbol\theta})$ and ${\mathbf B}(\bar{\boldsymbol\theta})$, respectively. The matrix-vector version can be simplified significantly by utilizing the special structure of the elements of $\bar{{\mathbf A}}(t)\in{\mathbb C}^{N\times L}$ and $\bar{{\mathbf B}}(t)\in{\mathbb C}^{N\times L}$. We now provide the details. Note that the $(n,l)$th element of $\bar{{\mathbf A}}(t)\in{\mathbb C}^{N\times L}$ and $\bar{{\mathbf B}}(t)\in{\mathbb C}^{N\times L}$ are ${\rm e}^{{\rm j}(n-(N-1)/2)\theta_l}/\sqrt{N}$ and ${\rm j}(n-(N-1)/2){\rm e}^{{\rm j}(n-(N-1)/2)\theta_l}/N^{\frac{3}{2}}$, where a symmetric index $n-(N-1)/2$ for $n=0,1,\cdots,N-1$ is used. Besides, $|\hat{a}_{nl}|^2$ can be simplified as 
\begin{align}
	|\hat{a}_{nl}|^2=|a_{nl}(\bar{\theta}_l)|^2+|b_{nl}(\bar{\theta}_l)|^2\hat{\epsilon}_l^2.	
\end{align}
Define ${\mathbf w}\in {\mathbb R}^{N}$ as 
\begin{align}
	w_n=\left(n-(N-1)/2\right)/N.	
\end{align}

Thus, $\bar{\boldsymbol{\nu}}^p(t)$ could be represented as 
\begin{align}
    \bar{\boldsymbol{\nu}}^p(t) & = |\hat{{\mathbf A}}(t)|^2 \boldsymbol{\nu}^x(t) + \boldsymbol{\nu}^a(t)|\hat{\mathbf{x}}(t)|^2\notag\\
    & =\frac{1}{N}{\mathbf 1}\left({\mathbf 1}^{\rm T}{\boldsymbol\nu}^x(t)\right)+\frac{1}{N}|{\mathbf w}|^2\left((|\hat{\boldsymbol \epsilon}(t)|^2)^{\rm T}{\boldsymbol\nu}^x(t)\right)\notag\\
    &+\frac{1}{N}|{\mathbf w}|^2\left((|\hat{{\mathbf x}}(t)|^2)^{\rm T}{\boldsymbol\nu}^{\epsilon}(t)\right)
\end{align}
where $|\hat{ {\mathbf A} }(t)|^2 = \frac{1}{N}\mathbf{1}_N\mathbf{1}_L^{\rm T} + \frac{1}{N}|\mathbf{w}|^2\left(|\hat{\epsilon}(t)|^2\right)^{\rm T}$ and $\boldsymbol{\nu}^a(t) = \frac{1}{N} (\boldsymbol{\nu}^\epsilon(t))^{\rm T} |\hat{\mathbf{x}}(t)|^2$.

$\boldsymbol{\nu}^p(t)$ can be simplified as
\begin{align}
	\boldsymbol{\nu}^p(t) = \bar{\boldsymbol{\nu}}^p(t) + \frac{1}{N}\left( (\boldsymbol{\nu}^x(t))^{\rm T}\boldsymbol{\nu}^\epsilon(t)\right)|\mathbf{w}|^2
\end{align}

Now we calculate the vector form of $\hat{\mathbf{p}}(t)$. We have
\begin{align}
	\hat{{\mathbf A}}(t)\hat{\mathbf{x}}(t) = \bar{\mathbf{A}}\hat{\mathbf{x}}(t) + {\rm j}\mathbf{w}\odot\left(\bar{\mathbf{A}}(\hat{\boldsymbol{\epsilon}}(t)\odot\hat{{\mathbf x}}(t))\right).
\end{align}
Utilizing the above equation, $\hat{\mathbf{p}}(t)$ can be computed as
\begin{small}
	\begin{flalign}
	\hat{\mathbf p}(t) =\bar{\mathbf A}\hat{\mathbf x}(t)+{\rm j}{\mathbf w}\odot\left(\bar{\mathbf A}\left(\hat{\mathbf x}(t)\odot\hat{\boldsymbol\epsilon}(t)\right)\right)-\hat{\mathbf s}(t)\odot\bar{\boldsymbol\nu}^p(t)
	\end{flalign}
\end{small}
Pesudo measurement and variance of $\mathbf{x}(t)$ can be simplified as
\begin{small}
	\begin{align}
		&\boldsymbol{\nu}^r(t) = \left( \left(\frac{1}{N}\mathbf{1}_L\mathbf{1}^{\rm T}_L + \frac{1}{N}|\hat{\boldsymbol{\epsilon}}(t)|^2(|\mathbf{w}|^2)^{\rm T}\right)\boldsymbol{\nu}^s(t) \right)^{-1}\notag\\
		&= \left(\frac{1}{N}(\mathbf{1}^{\rm T}_N\boldsymbol{\nu}^s(t))\mathbf{1}_L + \frac{1}{N}\left(\left(|\mathbf{w}|^2\right)^{\rm T}\boldsymbol{\nu}^s(t)\right)|\hat{\boldsymbol{\epsilon}}(t)|^2 \right)^{-1},
	\end{align}
\end{small}
\begin{align}
    \hat{\mathbf{r}}(t) &= \hat{\mathbf{x}}(t) + {\boldsymbol\nu}^r(t)\odot\left(\hat{\mathbf A}^{\rm H}\hat{\mathbf s}(t)-\hat{\mathbf x}(t)\odot\left(\left({\boldsymbol\nu}^a(t)\right)^{\rm T}|\hat{\mathbf s}(t)|^2\right)\right)\notag\\
    &= \hat{\mathbf{x}}(t) + {\boldsymbol\nu}^r(t)\odot\left(\bar{\mathbf{A}}^{\rm H}\hat{\mathbf{s}}(t)-{\rm j}\hat{\boldsymbol{\epsilon}}(t)\odot \left(\bar{\mathbf{A}}^{\rm H}\left(\mathbf{w}\odot \hat{\mathbf{s}}(t)\right)\right) \right.\notag\\
    & - \left.\frac{1}{N}\left((|\mathbf{w}|^2)^{\rm T}|\hat{\mathbf{s}}(t)|^2\right)\hat{\mathbf x}(t)\odot\boldsymbol{\nu}^\epsilon(t) \right),
\end{align}
where $\hat{\mathbf{A}}^{\rm H}(t)\hat{\mathbf{s}}(t) = \bar{\mathbf{A}}^{\rm H}\hat{\mathbf{s}}(t) -{\rm j} \hat{\boldsymbol{\epsilon}}(t)\odot \left(\bar{\mathbf{A}}^{\rm H}\left(\mathbf{w}\odot \hat{\mathbf{s}}(t)\right)\right)$, $\left(\boldsymbol{\nu}^a(t)\right)^{\rm T}|\hat{\mathbf{s}}(t)|^2 = \frac{1}{N}\left((|\mathbf{w}|^2)^{\rm T}|\hat{\mathbf{s}}(t)|^2\right)\boldsymbol{\nu}^\epsilon(t)$.

Finally, we simplify the vector from of $\boldsymbol{\nu}^q(t)$ and $\hat{\mathbf{q}}(t)$, that is
\begin{align}
    \boldsymbol{\nu}^q(t) &= \left(2|\hat{\mathbf x}(t)|^2\odot\left(|\bar{\mathbf B}^{\rm T}|^2{\boldsymbol\nu}^s(t)\right)\right)^{-1}\notag\\
    &= \left( \frac{2}{N}\left((|\mathbf{w}|^2)^{\rm T}\boldsymbol{\nu}^s(t)\right)|\hat{\mathbf{x}}(t)|^2 \right)^{-1},
\end{align}
where $\left(\bar{\mathbf{B}}^{\rm T}\right)^2\boldsymbol{\nu}^s(t) = \frac{1}{N}(|\mathbf{w}|^2)^T\boldsymbol{\nu}^s(t)\mathbf{1}_L$.
\begin{small}
    \begin{align}
    \hat{\mathbf q}(t) &= \hat{\boldsymbol\epsilon}(t) + {\boldsymbol\nu}^q(t)\odot 2\Re\left\{\hat{\mathbf x}^*(t)\odot \left(\bar{\mathbf B}^{\rm H}\hat{\mathbf s}(t)\right)\right\}\notag\\
    &= \hat{\boldsymbol{\epsilon}}(t)+ {\boldsymbol\nu}^q(t)\odot 2\Im\left\{\hat{\mathbf x}^*(t)\odot\left(\bar{\mathbf A}^{\rm H}(t)\left(\mathbf{w}\odot\hat{\mathbf{s}}(t)\right)\right)\right\},
\end{align}
\end{small}
where $\bar{\mathbf{B}}^{\rm H}\hat{\mathbf{s}}(t) = -{\rm j}\bar{\mathbf{A}}^{\rm H}(\mathbf{w}\odot \hat{\mathbf{s}}(t))$. 
\begin{algorithm}[htbp]
	\caption{BiG-LSE-Simplified}
	\begin{algorithmic}[1]
		\STATE Definition:\\
		\STATE $p_{z_n \mid p_n}\left(z \mid \hat{p} ; \nu^p\right)  \triangleq \frac{p_{y_n \mid z_n}\left(y \mid z\right) \mathcal{CN}\left(z;\hat{p}, \nu^p\right)}{\int_{z} p_{y_n \mid z_n}\left(y \mid z\right) \mathcal{CN}\left(z; \hat{p}, \nu^p\right){\rm d}z}$ \\
		\STATE $p_{x_l \mid r_l}\left(x \mid \hat{r}; \nu^r\right) \triangleq \frac{p_{x_l}(x) \mathcal{CN}\left(x ; \hat{r}, \nu^r\right)}{\int_{x} p_{x_l}\left(x\right) \mathcal{CN}\left(x ; \hat{r}, \nu^r\right){\rm d}x}$ \\
        \STATE $p_{\epsilon_l \mid q_l}\left(\epsilon \mid \hat{q} ; \nu^q\right) \triangleq \frac{p_{\mathrm{\epsilon}_l}(\epsilon) \mathcal{N}\left(\epsilon ; \hat{q}, \nu^q\right)}{\int_{\epsilon} p_{\mathrm{\epsilon}_l}\left(\epsilon\right) \mathcal{N}\left(\epsilon ; \hat{q}, \nu^q\right){\rm d}\epsilon}$\\
		\STATE Initialization:\\
		\STATE $\forall n: \hat{\mathbf s}(0)={\mathbf 0} $ \\
		\STATE $\forall l: \rm{set}\ \hat{\mathbf x}={\mathbf 0}_L,{\boldsymbol \nu}^x(1)={\rm var}_{\rm max}{\mathbf 1}_L,$\\
        \STATE $\hat{\boldsymbol \epsilon}(1)={\mathbf 0}_L,{\boldsymbol \nu}^\epsilon(1)={\rm var}_{\rm max}{\mathbf 1}_L$.
		\FOR{$t=1, \ldots T_{\max }$}
		\STATE $\hat{\mathbf A}(t) =  \bar{{\mathbf A}} +\bar{{\mathbf A}}\odot\left(\mathbf{w}\left(\hat{\boldsymbol{\epsilon}}(t)\right)^{\rm T}\right)$\\
		\STATE ${\boldsymbol\nu}^a(t) = \frac{1}{N}|\mathbf{w}|^2\left(\boldsymbol{\nu}^{\epsilon}(t)\right)^{\rm T} $ \\
		\STATE $\bar{\boldsymbol\nu}^p(t) =\frac{1}{N}{\mathbf 1}_{\rm N}\left({\mathbf 1}_{\rm L}^{\rm T}{\boldsymbol\nu}^x(t)\right)+\frac{1}{N}|{\mathbf w}|^2\left((|\hat{\boldsymbol \epsilon}(t)|^2)^{\rm T}{\boldsymbol\nu}^x(t)\right)$\\
        \STATE $+\frac{1}{N}|{\mathbf w}|^2\left((|\hat{{\mathbf x}}(t)|^2)^{\rm T}{\boldsymbol\nu}^{\epsilon}(t)\right)$\\
		\STATE ${\boldsymbol\nu}^p(t) = \bar{\boldsymbol\nu}^p(t) +\frac{1}{N}|{\mathbf w}|^2\left(({\boldsymbol\nu}^{x}(t))^{\rm T}{\boldsymbol\nu}^{\epsilon}(t)\right)$\\
		\STATE $ \hat{\mathbf p}(t) =\bar{\mathbf A}\hat{\mathbf x}(t)+{\mathbf w}\odot\left(\bar{\mathbf A}\left(\hat{\mathbf x}(t)\odot\hat{\boldsymbol\epsilon}(t)\right)\right)$ \\
        \STATE $-\hat{\mathbf s}(t)\odot\bar{{\boldsymbol\nu}}^p(t)$ \\
		\STATE ${\boldsymbol\nu}^z(t)={\mathrm {Var}}\left\{{\mathbf z} \mid {\mathbf p}=\hat{\mathbf p}(t) ; {\boldsymbol\nu}^p(t)\right\}$\\
		\STATE $\hat{\mathbf z}(t)={\mathrm E}\left\{{\mathbf z} \mid {\mathbf p}=\hat{\mathbf p}(t) ; {\boldsymbol\nu}^p(t)\right\}$\\
		\STATE ${\boldsymbol\nu}^s(t)=\left(1-{\boldsymbol\nu}^z(t) / {\boldsymbol\nu}^p(t)\right) / {\boldsymbol\nu}^p(t)$\\
		\STATE $\hat{\mathbf s}(t)=\left(\hat{\mathbf z}(t)-\hat{\mathbf p}(t)\right) / {\boldsymbol\nu}^p(t)$\\
		\STATE ${\boldsymbol\nu}^r(t)=\left(\frac{1}{N}(\mathbf{1}^{\rm T}_N\boldsymbol{\nu}^s(t))\mathbf{1}_L \right.$ \\
        \STATE $\left. + \frac{1}{N}\left(\left(|\mathbf{w}|^2\right)^{\rm T}\boldsymbol{\nu}^s(t)\right)|\hat{\boldsymbol{\epsilon}}(t)|^2 \right)^{-1}$\\
		\STATE $\hat{\mathbf r}(t)=\hat{\mathbf{x}}(t) + {\boldsymbol\nu}^r(t)\odot\left(\bar{\mathbf{A}}^{\rm H}\hat{\mathbf{s}}(t) + \hat{\boldsymbol{\epsilon}}(t)\odot \left(\bar{\mathbf{A}}^{\rm H}(\mathbf{w}^*\odot \hat{\mathbf{s}}(t))\right)\right)$\\
            \STATE $- {\boldsymbol\nu}^r(t)\odot\left(\frac{1}{N}\left((|\mathbf{w}|^2)^{\rm T}|\hat{\mathbf{s}}(t)|^2\right)\hat{\mathbf x}(t)\odot\boldsymbol{\nu}^\epsilon(t) \right)$\\
		\STATE ${\boldsymbol\nu}^q(t) = \left( \frac{2}{N}\left((|\mathbf{w}|^2)^{\rm T}\boldsymbol{\nu}^s(t)\right)|\hat{\mathbf{x}}(t)|^2 \right)^{-1}$\\
		\STATE $\hat{\mathbf q}(t)=\hat{\boldsymbol{\epsilon}}(t) + {\boldsymbol\nu}^q(t)\odot 2\Re\left\{\left(\hat{\mathbf x}(t)\odot\hat{\boldsymbol{\epsilon}}(t) \right)^*\odot\left(\bar{\mathbf A}^{\rm H}\left(\mathbf{w}^*\odot\hat{\mathbf{s}}(t)\right)\right)\right\}$\\
		\STATE ${\boldsymbol\nu}^x(t+1)=\mathrm{Var}\left\{{\mathbf x} \mid {\mathbf r}=\hat{\mathbf r}(t) ; {\boldsymbol\nu}^r(t)\right\}$\\
		\STATE $\hat{\mathbf x}(t+1)=\mathrm{E}\left\{{\mathbf x} \mid {\mathbf r}=\hat{\mathbf r}(t) ; {\boldsymbol\nu}^r(t)\right\}$\\
		\STATE ${\boldsymbol \nu}^\epsilon(t+1)=\mathrm{Var}\left\{{\boldsymbol\epsilon} \mid {\mathbf q}=\hat{\mathbf q}(t) ; {\boldsymbol\nu}^q(t)\right\}$\\
		\STATE $\hat{\boldsymbol\epsilon}(t+1)=\mathrm{E}\left\{{\boldsymbol\epsilon} \mid {\mathbf q}=\hat{\mathbf q}(t) ; {\boldsymbol\nu}^q(t)\right\}$\\
		\ENDFOR
	\end{algorithmic}
\end{algorithm}

\subsection{Initialization}\label{Initsubsec}
\begin{figure*}[htbp]
    \centering
    \includegraphics[width=\linewidth]{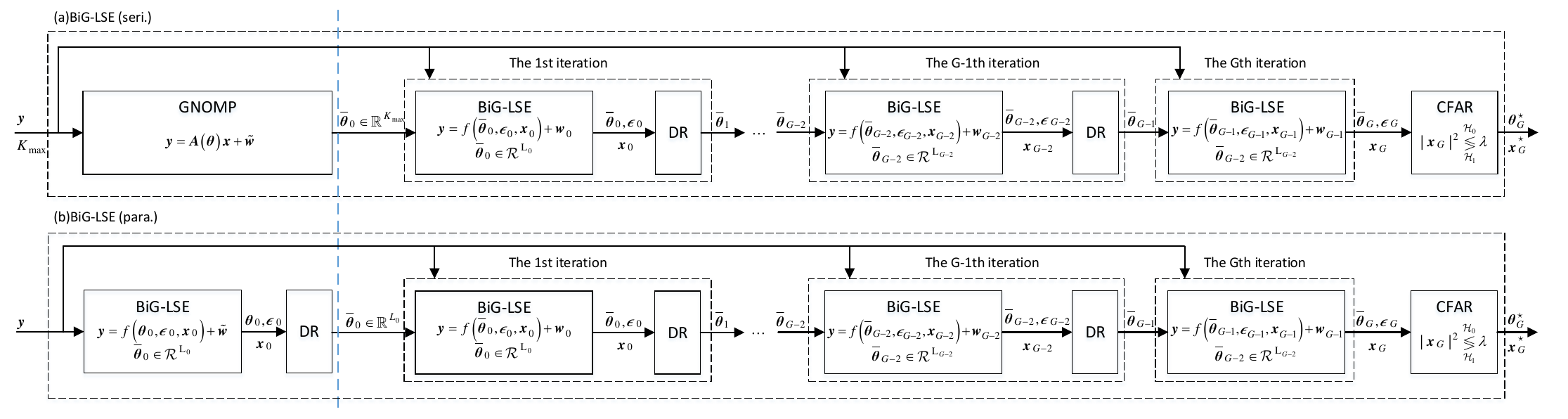}
    \caption{Two flowcharts of BiG-LSE: (a) BiG-LSE (para.), (b) BiG-LSE (seri.)}
    \label{frameflow}
\end{figure*}

This subsection presents two methods for obtaining $\bar{\boldsymbol\theta}$. The first is to discrete the frequency into a finite number of grids. Obviously, a natural strategy is to perform oversampling with respect to the Nyquist grid, i.e., the grid is constructed as ${\Theta}=\{0,2\pi/(\gamma_{\rm os} N),\cdots,,2\pi(\gamma_{\rm os} N-1)/(\gamma_{\rm os} N)\}$ with $\gamma_{\rm os}$ being the oversampling factor and we set $L=\gamma_{\rm os} N$. Therefore $ {\boldsymbol\epsilon}\in [-\pi/\gamma_{\rm os},\pi/\gamma_{\rm os}]^K$ and the variance of $\epsilon_l$ assuming a uniform distribution is $\frac{1}{12}\left(\frac{\pi}{\gamma_{\rm os}}\right)^2$. In the algorithm, $\epsilon_l$ is supposed to follow a Gaussian distribution and its variance is set as $\sigma_{\epsilon}^2=\frac{1}{12}\left(\frac{\pi}{\gamma_{\rm os}}\right)^2$. Note that the matrix operation $\bar{\mathbf A}(\cdot)$ and $\bar{\mathbf A}^{\rm H}(\cdot)$ can be implemented efficiently via the inverse fast Fourier transform (IFFT) and FFT. It is worth noting that a trade-off exists for $\gamma_{\rm os}$. For $\gamma_{\rm os}$ being small, the error in (\ref{zLsTaylor}) may be large and the estimation accuracy of $\boldsymbol\epsilon$ is not high. While for $\gamma_{\rm os}$ being large, the estimation accuracy of $\boldsymbol\epsilon$ should be high, but BiG-LSE is likely to be unstable due to the highly correlated structure. In our setting, we set $\gamma_{\rm os}=3$. The initialization can be viewed as a parallel method and the proposed algorithm is termed as \emph{\bf BiG-LSE (para.)}. The second is to use a greedy approach to obtain $\bar{\boldsymbol\theta}$ sequentially. The idea is to use the NOMP to obtain $\bar{\boldsymbol\theta}$, where a single Newton refinement is used to alleviate the off-grid errors. The oversampling in the NOMP is $4$. Note that all the single refinements, the cyclic refinements and the least squares steps in the NOMP are removed. The initialization can be viewed as a serial method and the proposed algorithm is termed as \emph{\bf BiG-LSE (seri.)}. 
A simple flowchart illustrating the entire process of the BiG-LSE framework is presented in Fig. \ref{frameflow}. In this flowchart, different initialization modules are selected to provide initial points based on BiG-LSE (seri.) and BiG-LSE (para.), where $f(\bar{\boldsymbol{\theta}},\boldsymbol{\epsilon},\mathbf{x}) = (\mathbf{A}(\bar{\boldsymbol{\theta}}) + \mathbf{B}(\bar{\boldsymbol{\theta}}){\rm diag}(\boldsymbol{\epsilon}))\mathbf{x}$. After initialization, the initial estimate $\bar{\boldsymbol{\theta}}_0$ is obtained and sent to the BiG-LSE module to perform the first iteration. The DR module receives $\bar{\boldsymbol{\theta}}_0$ and bias $\hat{\boldsymbol{\epsilon}}_0$ to correct the frequencies, i.e., $\bar{\boldsymbol{\theta}}_0 = \bar{\boldsymbol{\theta}}_0 + \frac{\hat{\boldsymbol{\epsilon}}_0}{N}$. The DR module then merges frequencies $\bar{\boldsymbol{\theta}}_0$ whose wrap around distances are smaller than the specified threshold, and outputs the frequency estimate $\bar{\boldsymbol{\theta}}_1$ for the next iteration. This process is repeated for a specified number $G$ of iterations, after which the estimates $\bar{\boldsymbol{\theta}}_G$ and $\hat{\mathbf{x}}_G$ are obtained and sent to the CFAR detector module. The CFAR detector with a threshold $\lambda$ computed by $-\hat{\sigma}^2\ln{P_{\rm FA}}$ \cite{CFAR_det} where $\hat{\sigma}^2$ and $P_{\rm FA}$ denote the noise variance estimate and probability of false alarm $P_{\rm FA}$, is aimed to further suppress false alarms and generate the final estimates.

 \subsection{Extension to the BiG-LSE (von-Mises)}

The previously developed BiG-LSE approximates the frequency $\theta_l$ via 
\begin{align}\label{epsilondef}
    \theta_l=\bar{\theta}_l+\epsilon_l/N,
\end{align}
where $\bar{\theta}_l$ is a known frequency obtained either by frequency discretezation or initialized by NOMP, $\epsilon_l/N$ is supposed to follow a Gaussian distribution $\epsilon_l/N\sim {\mathcal{N}}(0,\sigma_{\epsilon}^2/N^2)$ and the variance $\sigma_{\epsilon}^2/N^2$ is very small. Therefore $\theta_l$ follows a Gaussian distribution $\theta_l\sim {\mathcal{N}}(\bar{\theta}_l,\sigma_{\epsilon}^2/N^2)$, According to \cite{Direc}, for the Gaussian distribution whose variance is small, it can be approximated as a von Mises distribution, i.e., 
\begin{align}\label{priorofthetal}
    \theta_l\sim {\mathcal{VM}}(\bar{\theta}_l,\kappa_l),
\end{align}
where $\kappa_l$ is defined as $\kappa_l=N^2/\sigma_{\epsilon}^2$. In the following, we develop the von-Mises version of the BiG-LSE which treats the frequency $\{\theta_l\}_{l=1}^L$ as random variables instead of $\{\epsilon_l\}_{l=1}^L$. As a consequence, the prior distribution (\ref{priorofthetal}) is adopted. In the following, we replace these expressions involved with $\epsilon_l$ with $\theta_l$.
For line 9 and 10 in Algorithm \textbf{BiGLSE (scalar form)}, $\hat{a}_{nl}(t)$ is the expectation of ${a}_{nl}$ in the $t$th iteration taken with respect to the the posterior PDF of $\theta_l$ in the $(t-1)$th iteration. Let ${\mathcal VM}(\hat{\theta}_l,\hat{\kappa}_{l})$ denote the posterior PDF of $\theta_l$ in the $(t-1)$th iteration. $\hat{a}_{nl}(t)$ is 
\begin{align}
\hat{a}_{nl}(t)={\rm E}\left[{\rm e}^{{\rm j}n\theta_l}\right]=\frac{I_n(\hat{\kappa}_l)}{I_0(\hat{\kappa}_l)}{\rm e}^{{\rm j}n\hat{\theta}_l}.
\end{align}
$\nu^a_{nl}(t)$ denotes the posterior variance of $a_{nl}$ in the $t$th iteration taken with respect to the posterior PDF of $\theta_l$ in the $(t-1)$th iteration. It can be calculated that
\begin{align}
    \nu^a_{nl}(t) = \frac{1}{N}\left(1-\frac{I^2_{n}(\kappa^\theta_l)}{I^2_{0}(\kappa^\theta_l)}\right).
\end{align}
For line 20 and 21 in Algorithm \textbf{BiGLSE (scalar form)}, $\hat{q}_l$ and $\nu^q_l$ denote the pseudo observation and the variance of $\epsilon_l$. Utilize the fact  $b_{nl}(\bar{\theta_l}) = {\rm E}\left\{ \frac{\partial a_{nl}(\theta_l)}{\partial\theta_l} \right\}_{\theta_l\sim\mathcal{VM}(\hat{\theta}_l,\kappa^{\theta}_l) } = jn{\rm E}\left\{\frac{1}{\sqrt{N}}{\rm e}^{jn\theta_l}\right\}_{\theta_l\sim\mathcal{VM}(\hat{\theta}_l,\kappa^{\theta}_l) }$ $ = jn\hat{a}_{nl}(\theta_l)$. According to the definition of $\theta_l$ (\ref{epsilondef}),  
the pseudo observation and the variance of $\theta_l$ are 
\begin{small}
    \begin{align}
    &\hat{q}_{\theta_l} = \bar{\theta}_l +\frac{1}{N}\hat{\epsilon}_l + \frac{1}{N}\left(\nu^q_l\sum\limits^{N}_{n=1}2{\Re}\left\{\hat{s}_n \hat{b}_{nl}(\bar{\theta}_l)\hat{x}^\star_l\right\}\right)\notag\\
    &=\hat{\theta}_l(t) + N(\kappa_l^q(t))^{-1}\sum\limits^N_{n=1}2n{\Im}\left\{\hat{s}_n(t)\hat{a}_{nl}(t)\hat{x}^*_l(t)\right\},
    \label{ql_vm}\\
    &\nu^q_{\theta_l} = \frac{1}{N^2}\left(2\sum\limits^N_n|b_{nl}(\bar{\theta}_l)\hat{x}_l|^2\nu^s_n\right)^{-1}.
\end{align}
\end{small}
Because $\nu^q_{\theta_l}$ is usually very small, the Gaussian PDF is approximated as a von Mises PDF shown as 
\begin{align}
    \theta_l\sim {\mathcal{VM}}(\hat{q}_{\theta_l},\kappa_l^q),
\end{align}
where $\kappa_l^q=1/\nu^q_{\theta_l}$.

Finally, we assume the prior distribution of the frequency follows a von Mises distribution with a central direction \(\theta^p_l\) and concentration parameter \(\kappa^p_l\), i.e. $p(\theta_l)\sim \mathcal{VM}(\theta^p_l,\kappa^p_l)$. The pseudo-observation of \(\theta_l\) and the pseudo-variance, i.e. $\hat{q}_l$ and $\nu^q_l$, can be derived from that of $\epsilon_l$. Because the von Mises distribution is closed under multiplication, the posterior distribution of \(\theta_l\) is also von Mises distribution $\mathcal{VM}(\hat{\theta}_l, \kappa^{\theta}_l)$, and $(\hat{\theta}_l, \kappa^{\theta}_l)$ can be calculated as  
\begin{equation}
    \kappa^{\theta}_l e^{j\hat{\theta}_l} = \kappa^{q}_l e^{j\hat{q}_{\theta_l}} + \kappa^{p}_l e^{j\theta^p_l}.
\end{equation} 
Once multiple iterations are executed, we update the prior distribution by replacing their means with the posterior estimates of the frequencies. In this way, we iteratively refine the estimates and improve the estimation accuracy. 
\begin{algorithm}[htbp]
	\caption{BiG-LSE (VM)}
    	\begin{algorithmic}[1]\label{BiGLSEvm}
		\STATE Definition:\\
		\STATE $p_{z_{n} \mid p_{n}}\left(z \mid \hat{p} ; \nu^p\right)  \triangleq \frac{p_{y_{n} \mid z_{n}}\left(y \mid z\right) \mathcal{CN}\left(z;\hat{p}, \nu^p\right)}{\int_{z_{n}} p_{y_{n} \mid z_{n}}\left(y \mid z\right) \mathcal{CN}\left(z;\hat{p}, \nu^p\right){\rm d}z}$ \\
		\STATE $p_{x_l \mid r_l}\left(x \mid \hat{r} ; \nu^r\right) \triangleq \frac{p_{x_l}(x) \mathcal{CN}\left(x ; \hat{r}, \nu^r\right)}{\int_{x} p_{x_l}\left(x\right) \mathcal{CN}\left(x ; \hat{r}, \nu^r\right){\rm d}x}$ \\
		\STATE $p_{\theta_l \mid q_{\theta_l}}\left(\theta \mid \hat{q}_{\theta} ; \kappa^{q_{\theta}}\right) \triangleq \frac{p_{\mathrm{\theta}_l}(\theta) \mathcal{VM}\left(\theta ; \hat{q}_{\theta}, \kappa^{q_{\theta}}\right)}{\int_{\theta} p_{\theta_l}\left(\theta\right) \mathcal{VM}\left(\theta ; \hat{q}_{\theta}, \kappa^{q_\theta}\right){\rm d}\theta}$\\
		\STATE Initialization:\\
		\STATE $\forall n: \hat{s}_{n}(0)=0 $ \\
		\STATE $\forall l: \rm{set}\ \hat{\theta}_{l}(1),\kappa^{\theta}_{l}(1),\hat{x}_l(1),\nu^x_l(1)$
		\FOR{$t=1, \ldots T_{\max }$}
		\STATE $\forall n,l: \hat{a}_{nl}(t) =  a_{nl}(\hat{\theta}_l)\frac{I_n(\kappa^\theta_l)}{I_0(\kappa^\theta_l)}$\\
		\STATE $\forall n,l: \nu^a_{nl}(t) = \frac{1}{N}\left(1-\frac{I^2_{n}(\kappa^\theta_l)}{I^2_{0}(\kappa^\theta_l)}\right)$ \\
		\STATE $\forall n: \bar{\nu}^p_n(t) = \sum_{l}|\hat{a}_{nl}(t)|^2 \nu^x_l(t) + \sum_{l} \nu^a_{nl}(t)|\hat{x}_l(t)|^2$\\
		\STATE $\forall n: \nu^p_n(t) = \bar{\nu}^p_n(t) + \sum_{l} \nu^x_l(t) \nu^a_{nl}(t) $\\
		\STATE $\forall n: \hat{p}_n(t) = \sum_{l}\hat{a}_{nl}(t)\hat{x}_l(t)-\hat{s}_n(t-1)\bar{\nu}^p_n(t) $\\
		\STATE $\forall n: \nu^z_n(t)={\mathrm {Var}}\left\{{z}_{n} \mid {p}_{n}=\hat{p}_{n}(t) ; \nu_{n}^p(t)\right\}$\\
		\STATE $\forall n: \hat{z}_{n}(t)={\mathrm E}\left\{{z}_{n} \mid {p}_{n}=\hat{p}_{n}(t) ; \nu_{n}^p(t)\right\}$\\
		\STATE $\forall n: \nu_{n}^s(t)=\left(1-\nu_{n}^z(t) / \nu_{n}^p(t)\right) / \nu_{n}^p(t)$\\
		\STATE $\forall n: \hat{s}_{n}(t)=\left(\hat{z}_{n}(t)-\hat{p}_{n}(t)\right) / \nu_{n}^p(t)$\\
		\STATE $\forall l: \nu^r_l(t)=\left(\sum_{n}|\hat{a}_{nl}(t)|^2\nu_{n}^s(t) \right)^{-1}$\\
		\STATE $\forall l: \hat{r}_l(t)=\hat{x}_l(t) + \nu^r_l(t)\sum_{n}\hat{a}^*_{nl}(t)\hat{s}_n(t) - \nu^r_l(t)\hat{x}_l(t)\sum_n\nu^a_{nl}(t)|\hat{s}_n(t)|^2$\\ 
		\STATE $\forall l: \kappa_l^{q_\theta}(t) = 2N^2\sum_{n}n^2|\hat{x}_l(t)|^2 \nu_{n}^s(t)$\\
		\STATE $\forall l: \hat{q}_{\theta_l}(t)=\hat{\theta}_l(t) + N(\kappa_l^{q_\theta}(t))^{-1}\sum_n 2n{\Im}\left\{\hat{s}_n(t)\hat{a}_{nl}(t)\hat{x}^*_l(t)\right\}$\\
		\STATE $\forall l: \nu_l^x(t+1)=\mathrm{Var}\left\{x_l \mid r_l=\hat{r}_l(t) ; \nu_l^r(t)\right\}$\\
		\STATE $\forall l: \hat{x}_l(t+1)=\mathrm{E}\left\{x_l \mid r_l=\hat{r}_l(t) ; \nu_l^r(t)\right\}$\\
		\STATE $\forall l: \kappa^\theta_l(t+1)=\mathrm{Var}\left\{\theta_l \mid q_{\theta_l}=\hat{q}_{\theta_l}(t) ; \kappa_l^{q_\theta}(t)\right\}$\\
		\STATE $\forall l: \hat{\theta}_l(t+1)=\mathrm{E}\left\{\theta_l \mid q_{\theta_l}=\hat{q}_{\theta_l}(t) ; \kappa_l^{q_\theta}(t)\right\}$\\
		\ENDFOR
    \end{algorithmic}
\end{algorithm}

\section{Simulation and Experiment}
In this section, numerical simulations and real experiments are conducted to evaluate the performance of BiG-LSE (para.) and BiG-LSE (seri.). EM is used to estimate the nusiance parameters of the prior distribution \cite{Vila}. The codes implementing these versions are available \cite{BiGLSEsoftware}. The following benchmark algorithms are used to make performance comparison:
\begin{itemize}
	\item Superfast LSE \cite{HanseSuperfast}: A Bayesian algorithm achieves
	estimation accuracy at least as good as current methods while being orders of magnitudes faster.
	\item VALSE \cite{Badiu}: A Bayesian algorithm that provides uncertain degrees of the frequencies estimates.
	\item NOMP \cite{Mamandipoor}: A greedy algorithm that incorporates the Newton step into the OMP to achieve high estimation accuracy and maintain the CFAR detection.
	\item GLS \cite{GLS}: an ANM based approach to perform the line spectrum estimation. 
\end{itemize} 

We use three metrics, the normalized mean squared error (NMSE) ${\rm NMSE}(\hat{\mathbf z})$, the false alarm probability $P_{\rm FA}$ and the detection probability $P_{\rm D}$ to evaluate the signal estimation and detection performance. Specifically, the NMSE ${\rm NMSE}(\hat{\mathbf z})$ is defined as 
\begin{align}
	{\rm NMSE}(\hat{\mathbf z})=10\log\left({\|\hat{\mathbf z}-{\mathbf z}\|_2^2}/{\|{\mathbf z}\|_2^2}\right),
\end{align}
where the signal estimate $\hat{\mathbf z}$ is defined as $\hat{\mathbf z}\triangleq {\mathbf A}(\hat{\boldsymbol \theta}) \hat{\mathbf x}$, the frequencies estimates are $\hat{\boldsymbol\theta}=\bar{\boldsymbol \theta}+\hat{\boldsymbol\epsilon}/N$. In traditional radar system, the continuous frequency $\theta\in[0,2\pi]$ is often discretized into $N$ independent cells, and the false alarm probability $P_{\rm FA}$ is defined as
\begin{align}
	P_{\rm FA} = {K_{\rm FA}}/{({\rm MC} \times N)},
\end{align}
where ${\rm MC}$ denotes the number of Monte Carlo (MC) trials, $ K_{\rm FA}$ is the total number of false alarms. An estimated frequency $\hat{\omega}$ is selected as a false alarm if ${\rm min}_{k=1,\cdots,K} {\rm dist}(\hat{\omega},\omega_k)\geq 0.5\Delta\theta_{\rm min}$, where $K$ denotes the number of targets, i.e., the number of frequencies, and $\Delta \theta_{\rm min}$ denotes the minimal wrap around distance between any two generated frequencies in the simulation setup, i.e., their wrap around distance is greater than $\Delta \theta_{\rm min}$. The detection probability $P_{\rm D}$ is defined as
\begin{align}
	P_{\rm D} = {K_{\rm D}}/{{\rm MC}},
\end{align}
where $K_{\rm D}$ is the number of trials in which the weakest target is detected.  An frequency $\omega$ is detected if ${\rm min}_{k=1,\cdots,\hat{K}} {\rm dist}(\hat{\omega}_k,\omega)\leq 0.5\Delta \theta_{\rm min}$, where $\hat{K}$ denotes the estimated model order, i.e., the number of estimated frequencies. Additionally, we measure the average running time of each algorithm, plot the cumulative distribution function (CDF) curves of the model order estimation for different algorithms under various scenarios.

Simulation settings are as follows: $N = 1024$. The frequencies are generated randomly from $[0,2\pi)$ while satisfying the minimum separation constraints specified by $\Delta \theta_{\rm min}$. Four scenarios listed in Table \ref{TableScenario} are simulated. For $K$ targets, the integrated SNR of a target is varied from 8 dB to 20 dB, and the integrated SNR of the remaining targets are fixed as 22 dB.  All the results are averaged over $500$ MC trials.
\begin{table}[ht]
	\centering
	\caption{Scenario settings where $\Delta \theta_{\rm DFT}=2\pi/N$.}\label{TableScenario}
	\begin{tabular}{c|c|c}
		\hline
		Scenario ID & ${\Delta \theta_{\rm min}}/\Delta \theta_{\rm DFT}$ & Number of targets $K$ \\
		\hline
		Scenario 1 & 1 & 10 \\
		\hline
		Scenario 2 & 0.5 & 10 \\
		\hline
		Scenario 3 & 1 & 200 \\
		\hline
		Scenario 4 & 0.5 & 200 \\
		\hline
	\end{tabular}
\end{table}

\subsection{Scenario $1$ and Scenario $2$}
\begin{figure*}[htb]
	\centering
		\subfigure[]{
			\includegraphics[width=0.45\textwidth]{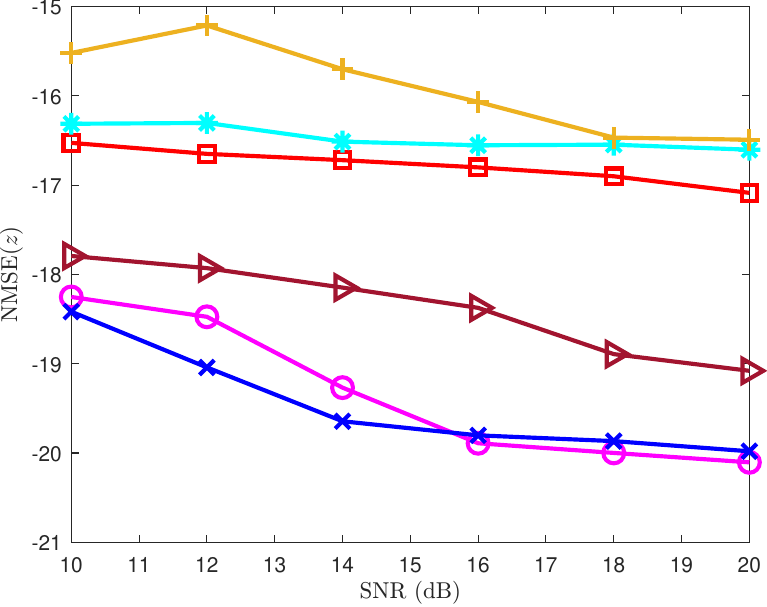}
			\label{NMSE_SNR_ly}}
		\subfigure[]{
			\includegraphics[width=0.45\textwidth]{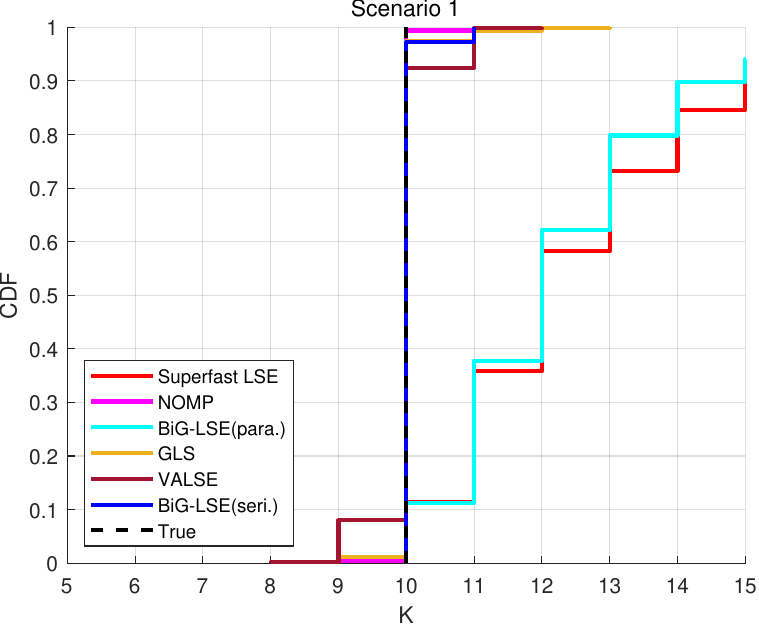}
			\label{Pro_K_ly}}
		\subfigure[]{
			\includegraphics[width=0.45\textwidth]{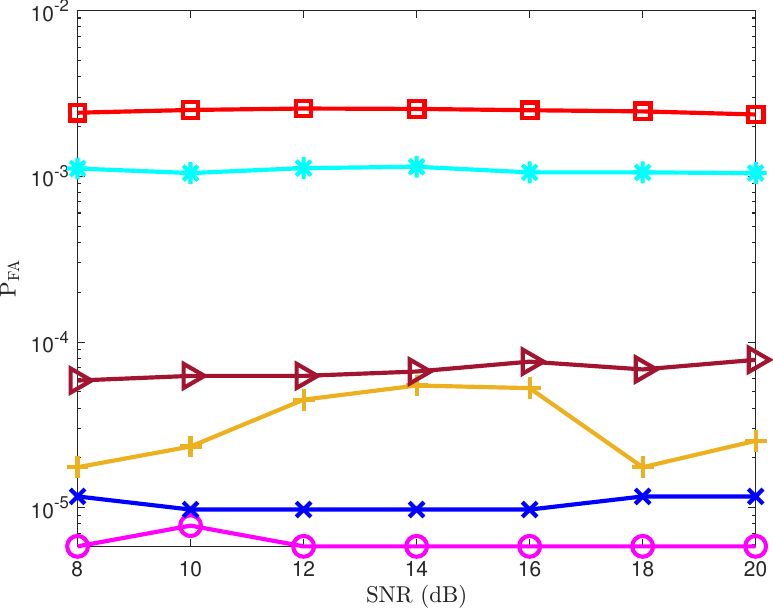}
			\label{Pfa_SNR_ly}}
		\subfigure[]{
			\includegraphics[width=0.45\textwidth]{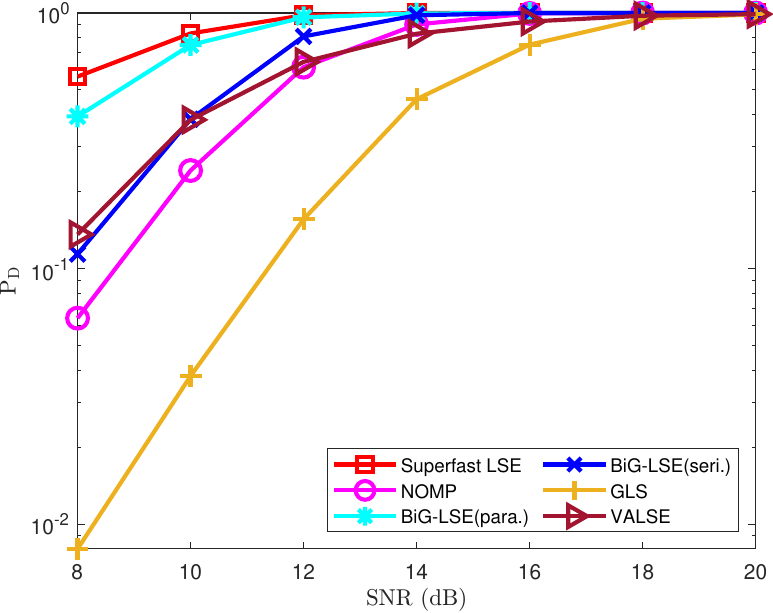}
			\label{Pd_SNR_ly}}
		\caption{The NMSE ${\rm NMSE}(\hat{\mathbf z})$, the CDF of the model order estimates, the false alarm probability $P_{\rm FA}$ and the detection probability $P_{\rm D}$ under Scenario $1$: (a) NMSE ${\rm NMSE}(\hat{\mathbf z})$ versus SNR, (b) the CDF of the model order estimation under ${\rm SNR}=$ 16 dB, (c) the false alarm probability $P_{\rm FA}$ versus SNR, (d) the detection probability $P_{\rm D}$ versus SNR.}\label{resultsScenario1}
\end{figure*}
\begin{figure*}[htbp]
	\centering
	\subfigure[]{
		\includegraphics[width=0.45\textwidth]{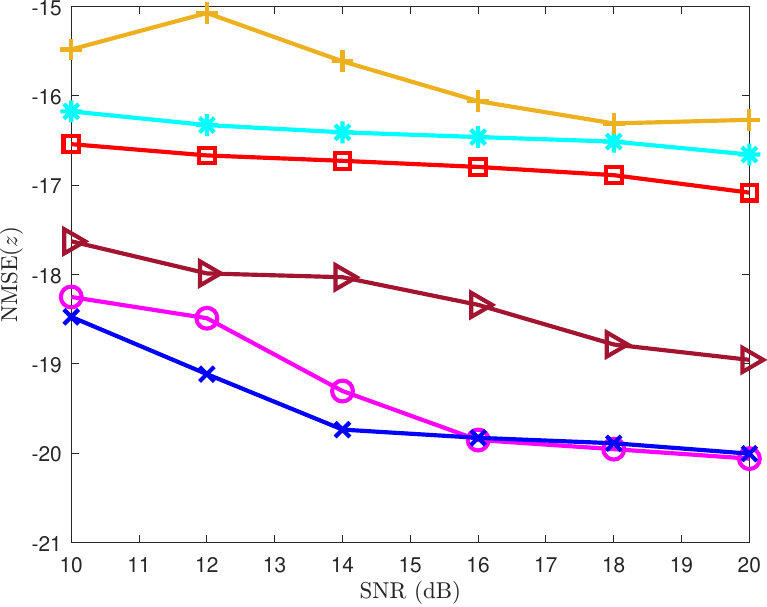}
		\label{NMSE_SNR_ln}}
	\subfigure[]{
		\includegraphics[width=0.45\textwidth]{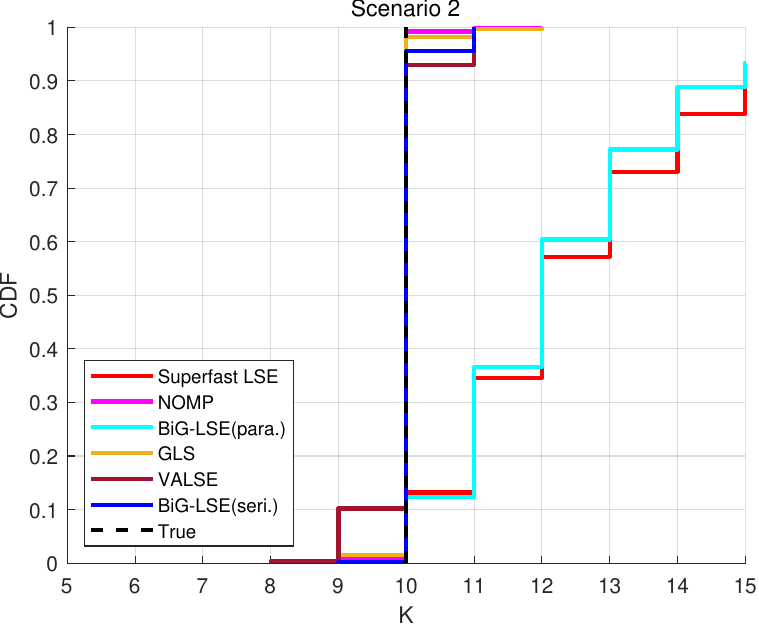}
		\label{Pro_K_ln}}
	\subfigure[]{
		\includegraphics[width=0.45\textwidth]{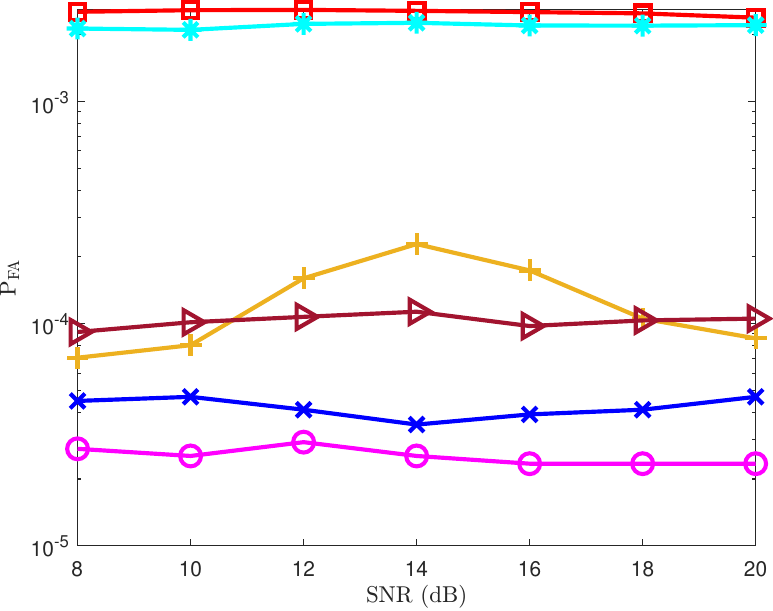}
		\label{Pfa_SNR_ln}}
	\subfigure[]{
		\includegraphics[width=0.45\textwidth]{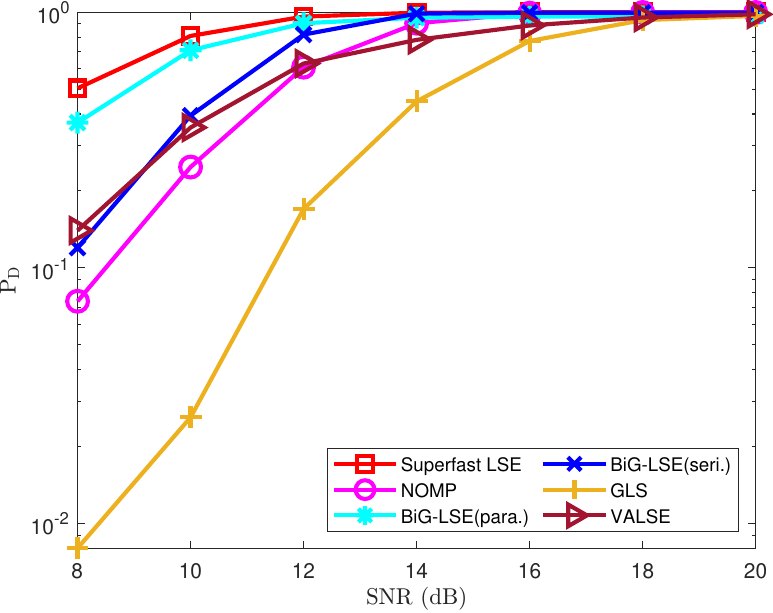}
		\label{Pd_SNR_ln}}
	\caption{The NMSE ${\rm NMSE}(\hat{\mathbf z})$, the CDF of the model order estimates, the false alarm probability $P_{\rm FA}$ and the detection probability $P_{\rm D}$ under Scenario $2$: (a) NMSE ${\rm NMSE}(\hat{\mathbf z})$ versus SNR, (b) the CDF of the model order estimates under ${\rm SNR}=$ 16 dB, (c) the false alarm probability $P_{\rm FA}$ versus SNR, (d) the detection probability $P_{\rm D}$ versus SNR.}\label{Resultsscenario2}
\end{figure*}

The NMSE, the cumulative distribution functions (CDFs) of the model order estimation, the false alarm probability and detection probability of all algorithms are plotted in Fig. \ref{resultsScenario1}. Fig. \ref{NMSE_SNR_ly} shows that BiG-LSE (seri.) performs best in terms of the signal estimation error, followed by the NOMP, VALSE, Superfast LSE, BiG-LSE (para.) and GLS. Fig. \ref{Pro_K_ly} shows that Superfast LSE and  BiG-LSE (para.) tend to overestimate the model order, while VALSE tends to underestimate it. Note that the means of the distributions of NOMP, BiG-LSE (seri.) and GLS are close to the truth (small bias), and they have a small spread around the mean (small variance), which means they provide more accurate model order estimates than the rest of algorithms. Compared to BiG-LSE (seri.) and GLS, NOMP has a higher probability of underestimating the model order, slightly above 0.1. As for the $P_{\rm FA}$ and $P_{\rm D}$, it can be seen that although Superfast LSE has a high detection probability, it also has a high false alarm probability. Compared to Superfast LSE, the false alarm probability and detection probability of BiG-LSE (para.) is slightly lower. Both Superfast LSE and BiG-LSE (para.) incorrectly estimate the model order, overestimating many spurious components, which can be confirmed by Fig. \ref{Pro_K_ly}. The false alarm probability of VALSE and GLS are higher than that of BiG-LSE (seri.), while their detection probabilities are lower. The detection performance of NOMP is close to that of BiG-LSE (seri.). Similar conclusion could be drawn from the results shown in Fig. \ref{Resultsscenario2} under scenario 2.

Moreover, we measured the running time of each algorithm in both scenarios shown in Table \ref{runtime1}. It can be seen that NOMP is the fastest, followed by Superfast LSE and BiG-LSE (seri.) which are on the same order, BiG-LSE (para.), VALSE and GLS. Note that both VALSE and GLS take hundreds of seconds even for a smaller number of targets, we do not evaluate their performance under Scenario 3 and Scenario 4 for large number of targets.
\begin{table*}[htbp]
	\centering
	\caption{Time (Sec) of each algorithm under Scenario 1 and Scenario 2}
	\begin{tabular}{c|c|c|c|c|c|c}
		\hline
		\diagbox{Scenario}{Algorithm} & Superfast LSE & NOMP & BiG-LSE (seri.) & BiG-LSE (para.) & VALSE & GLS \\
		\hline
		Scenario 1 & 0.12 & 0.04 & 0.67 & 9.30 & 135.45 & 970.30\\
		\hline
		Scenario 2 & 0.16 & 0.05 & 0.83 & 11.62 & 177.23 & 1308.60\\
		\hline
	\end{tabular}
	\label{runtime1}
\end{table*}
\subsection{Scenario $3$ and Scenario $4$}
\begin{figure*}[htbp]
	\centering
	\subfigure[]{
		\includegraphics[width=0.45\textwidth]{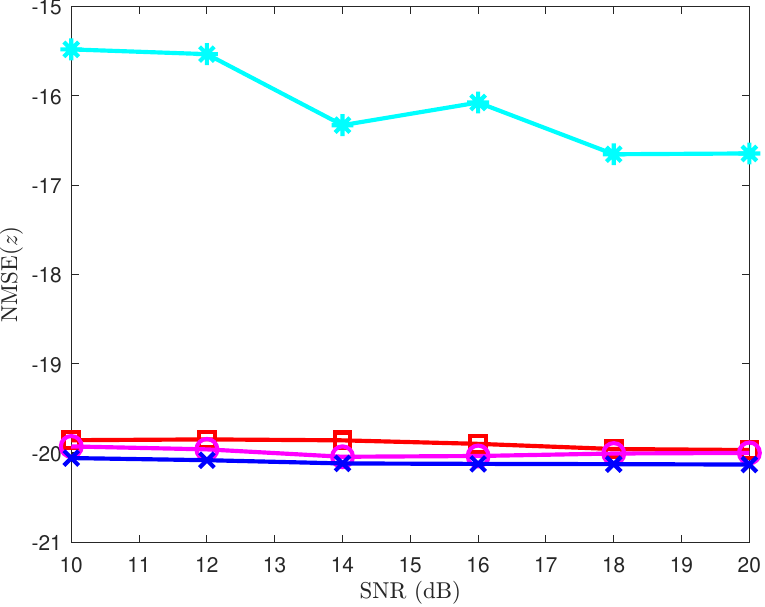}
		\label{NMSE_SNR_my}}
	\subfigure[]{
		\includegraphics[width=0.45\textwidth]{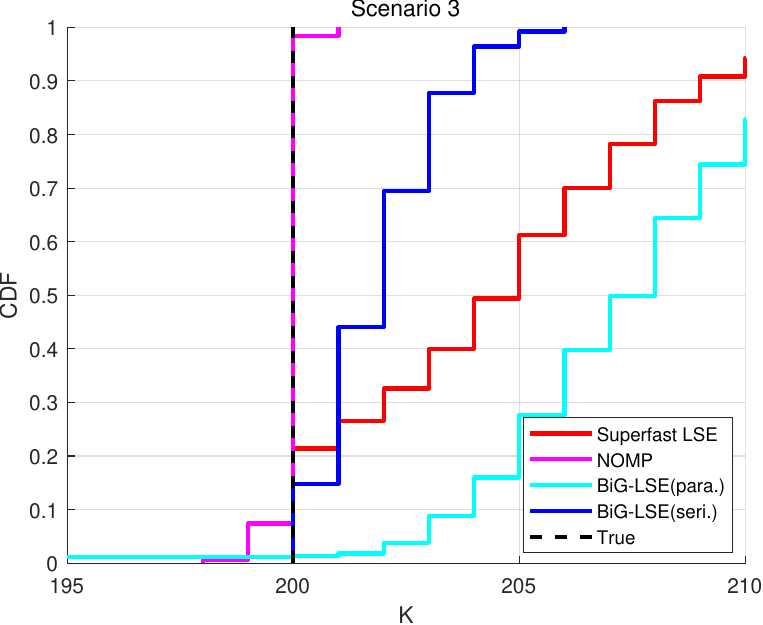}
		\label{Pro_K_my}}
	\subfigure[]{
		\includegraphics[width=0.45\textwidth]{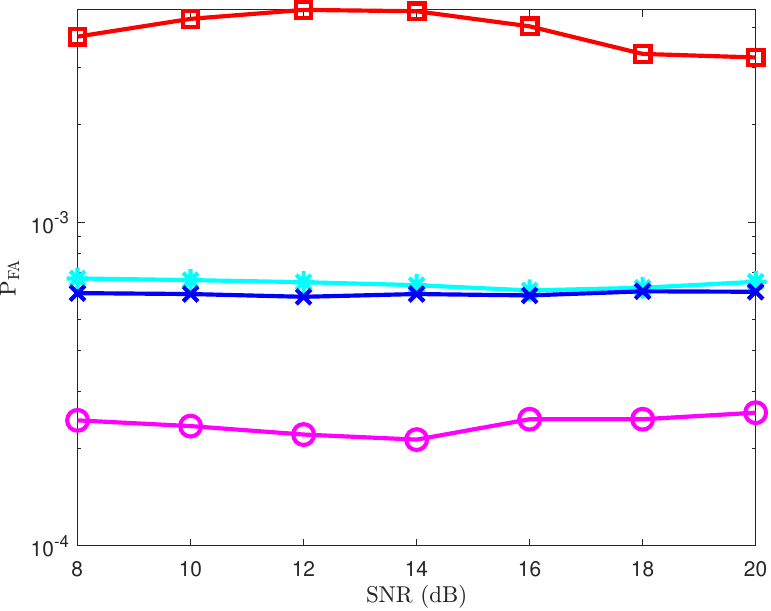}
		\label{Pfa_SNR_my}}
	\subfigure[]{
		\includegraphics[width=0.45\textwidth]{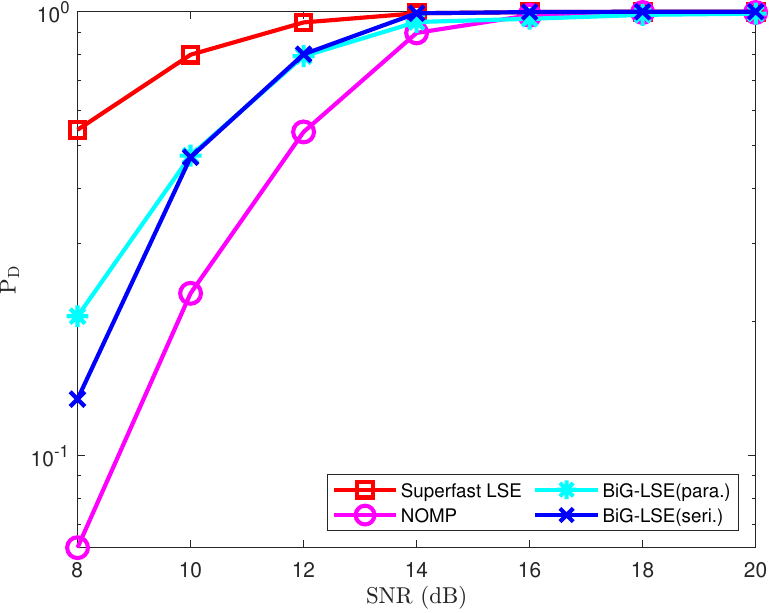}
		\label{Pd_SNR_my}}
	\caption{The NMSE ${\rm NMSE}(\hat{\mathbf z})$, the CDF of the model order estimates, the false alarm probability $P_{\rm FA}$ and the detection probability $P_{\rm D}$ under Scenario $3$: (a) NMSE ${\rm NMSE}(\hat{\mathbf z})$ versus SNR, (b) the CDF of the model order estimates under ${\rm SNR}=$ 16 dB, (c) the false alarm probability $P_{\rm FA}$ versus SNR, (d) the detection probability $P_{\rm D}$ versus SNR.} \label{Resultsscenario3}
\end{figure*}
 
Under the parameter settings of Scenarios 3, we plot the results shown in Fig. \ref{Resultsscenario3} . From Fig. \ref{NMSE_SNR_my}, it can be observed that the NMSE of all algorithms is insensitive to changes in SNR. This is because the number of targets is very large and only the SNR of the weakest target is varied. BiG-LSE (seri.) achieves the best performance, followed by NOMP and Superfast LSE, BiG-LSE (para.). Fig. \ref{Pro_K_my} shows that all algorithms, except for NOMP, tend to overestimate the model order. Compared to Superfast LSE and BiG-LSE (para.), NOMP and BiG-LSE (seri.) provide more accurate model order estimates. From Fig. \ref{Pd_SNR_my}, Superfast LSE has the highest detection probability, followed by BiG-LSE (para.), BiG-LSE(seri.) and NOMP, while Superfast LSE also has the highest false alarm probability, followed by BiG-LSE (para.), BiG-LSE (seri.) and NOMP, with the false alarm rate of BiG-LSE (seri.) being approximately 2.5 times than that of NOMP.
\begin{figure*}[htbp]
	\centering
	\subfigure[]{
		\includegraphics[width=0.45\textwidth]{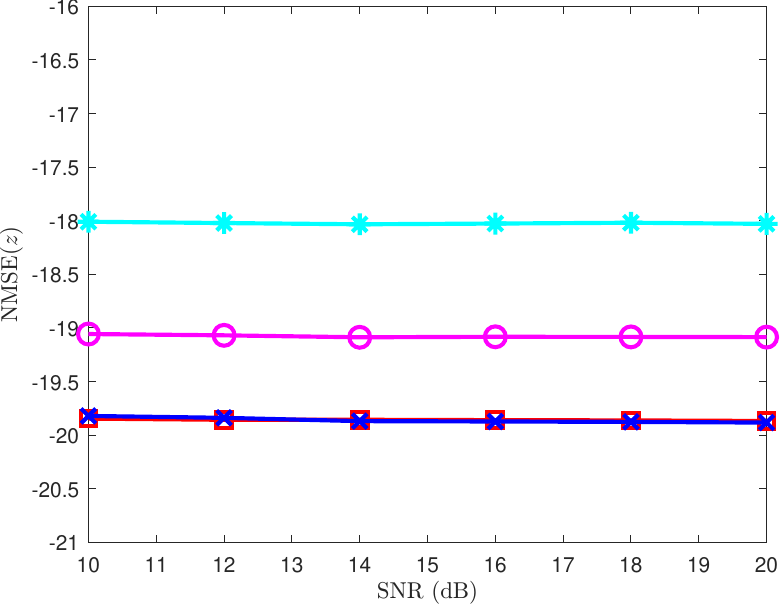}
		\label{NMSE_SNR_mn}}
	\subfigure[]{
		\includegraphics[width=0.45\textwidth]{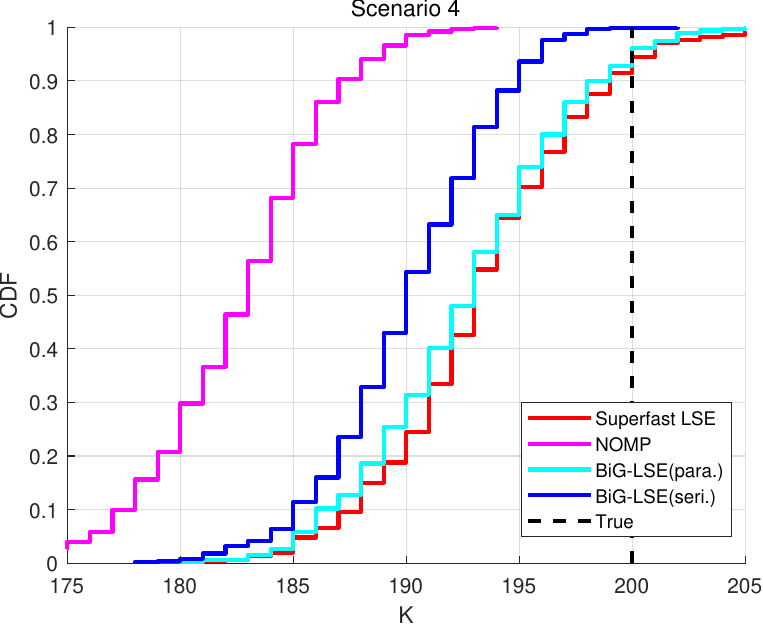}
		\label{Pro_K_mn}}
	\subfigure[]{
		\includegraphics[width=0.45\textwidth]{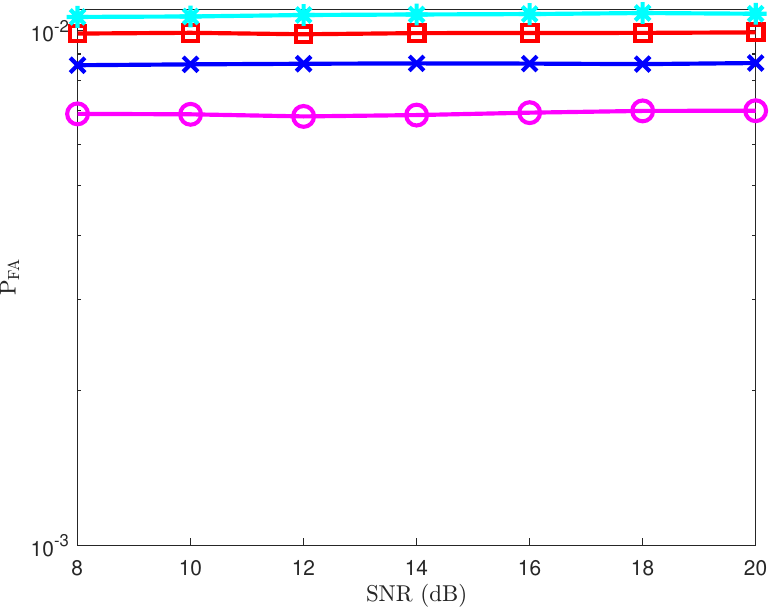}
		\label{Pfa_SNR_mn}}
	\subfigure[]{
		\includegraphics[width=0.45\textwidth]{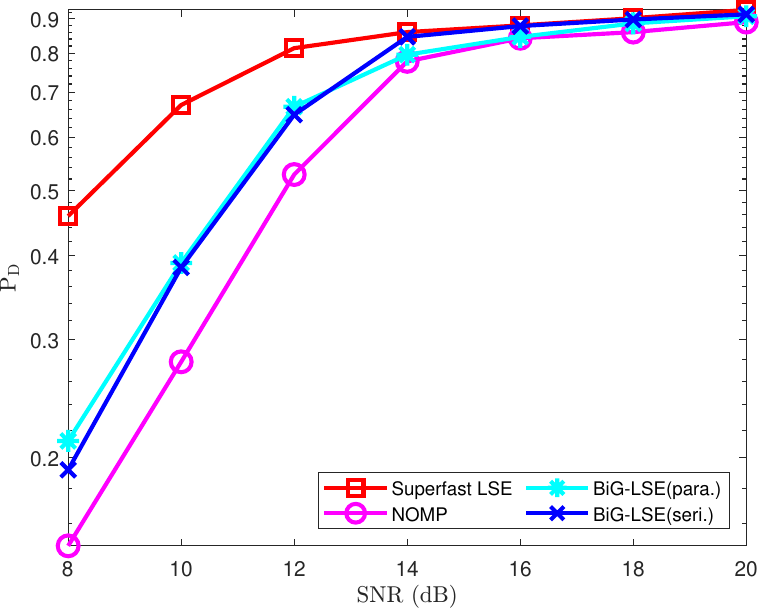}
		\label{Pd_SNR_mn}}
	\caption{The NMSE ${\rm NMSE}(\hat{\mathbf z})$, the CDF of the model order estimates, the false alarm probability $P_{\rm FA}$ and the detection probability $P_{\rm D}$ under Scenario $4$: (a) NMSE ${\rm NMSE}(\hat{\mathbf z})$ versus SNR, (b) the CDF of the model order estimates under ${\rm SNR}=$ 16 dB, (c) the false alarm probability $P_{\rm FA}$ versus SNR, (d) the detection probability $P_{\rm D}$ versus SNR.} \label{Resultsscenario4}
\end{figure*}
\begin{table*}[htbp]
	\centering
	\caption{Running times under scenario 3 and scenario 4, units/s}
	\begin{tabular}{c|c|c|c|c}
		\hline
		\diagbox{Scenario}{Algorithm} & Superfast LSE & NOMP & BiG-LSE (seri.) & BiG-LSE (para.) \\
		\hline
		Scenario 3 & 0.43 & 10.14 & 7.55 & 13.83 \\
		\hline
		Scenario 4 & 0.59 & 8.50 & 7.23 & 13.22 \\
		\hline
	\end{tabular}
	\label{runtime2}
\end{table*}
The NMSE, the CDFs of the model order estimation, the false alarm probability and detection probability of all algorithms under Scenario $4$ are plotted in Fig. \ref{Resultsscenario4}. Fig. \ref{NMSE_SNR_mn} shows that both BiG-LSE (seri.) and Superfast LSE perform best in terms of the signal estimation error, followed by the NOMP and BiG-LSE (para.). Fig. \ref{Pro_K_mn} shows that all the algorithms tend to underestimate the model order, i.e., the means of the distributions of all the algorithms are far from the truth. Compared with Scenario 3, NOMP degrades significantly in terms of the NMSE and model order estimate performances due to the decrease of $\Delta \theta_{\rm min}$ between targets. 
As for the $P_{\rm FA}$ and $P_{\rm D}$, it can be seen that Superfast LSE achieves the highest detection probability, followed by BiG-LSE (para.), BiG-LSE (seri.) and NOMP, while the false alarm probability of Superfast LSE is close to that of BiG-LSE (para.). The false alarm probability of NOMP is close to BiG-LSE (seri.), while the detection probability of NOMP is lower than that of BiG-LSE (seri.). 

The running time of each algorithm under scenarios 3 and scenario 4 are shown in Table \ref{runtime2}. SuperFast LSE has the fastest running time, followed by BiG-LSE (seri.), NOMP and BiG-LSE (para.). Compared with scenario 1 and 2, the running time of Superfast LSE almost keeps unchanged, and the running time of NOMP increases significantly and surpasses the BiG-LSE (seri.). 

In summary, it can be seen that the proposed BiG-LSE (seri.) almost achieves best in terms of signal reconstruction error under the four scenarios, and performs well in terms of $P_{\rm D}$, $P_{\rm FA}$ and model order estimation. which are close to the NOMP and better that the other algorithms. Besides, the proposed BiG-LSE (seri.) can be easily extended to handle nonlinear measurement scenarios \cite{mengunified}, as we show in the ensuing subsection.

\subsection{Real Experiment}
\begin{figure}[htbp]
	\centering
	\includegraphics[width=0.65\textwidth]{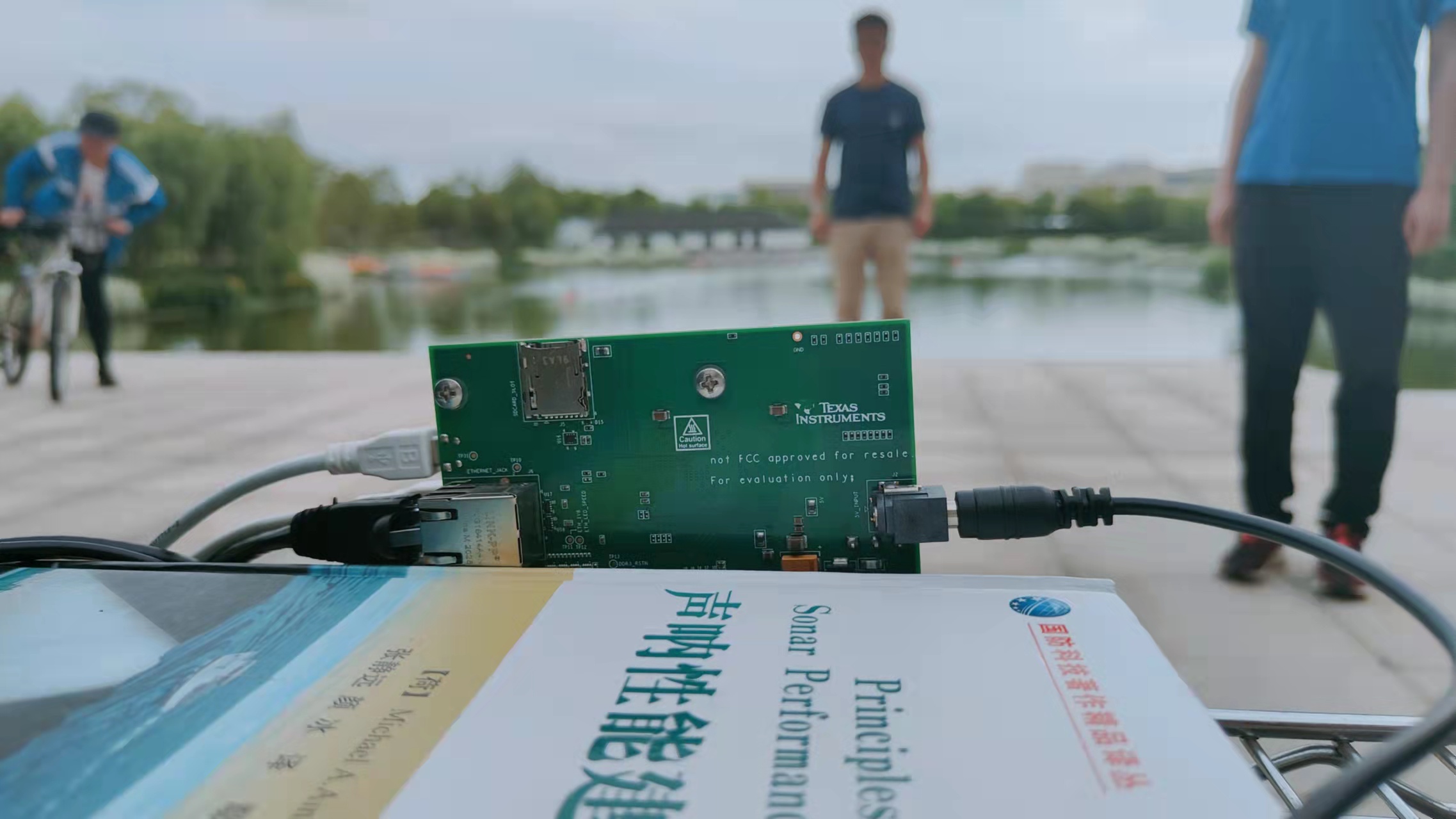}
	\caption{The setup of real experiment}
	\label{exp_real_2}
\end{figure}

\begin{figure*}[htbp]
	\centering
	\subfigure[]{
		\includegraphics[width=0.45\textwidth]{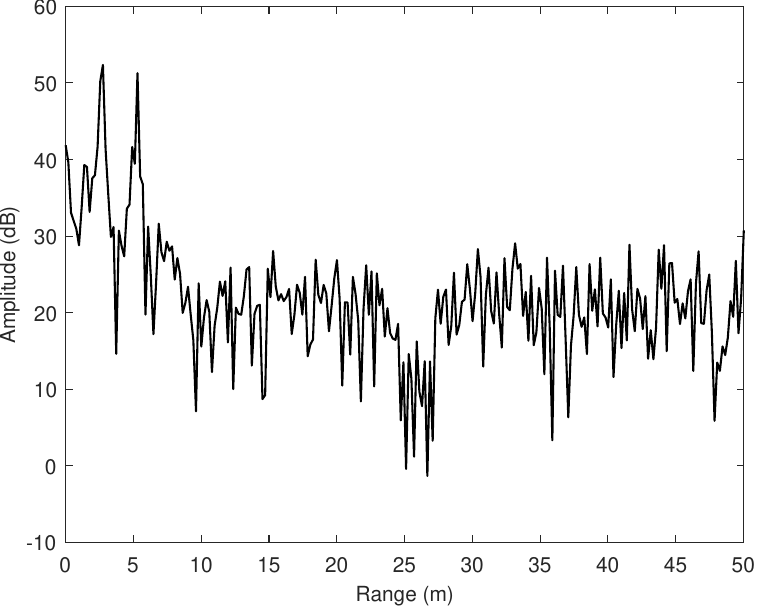}
		\label{exp_real_2_fft}}
	\subfigure[]{
		\includegraphics[width=0.45\textwidth]{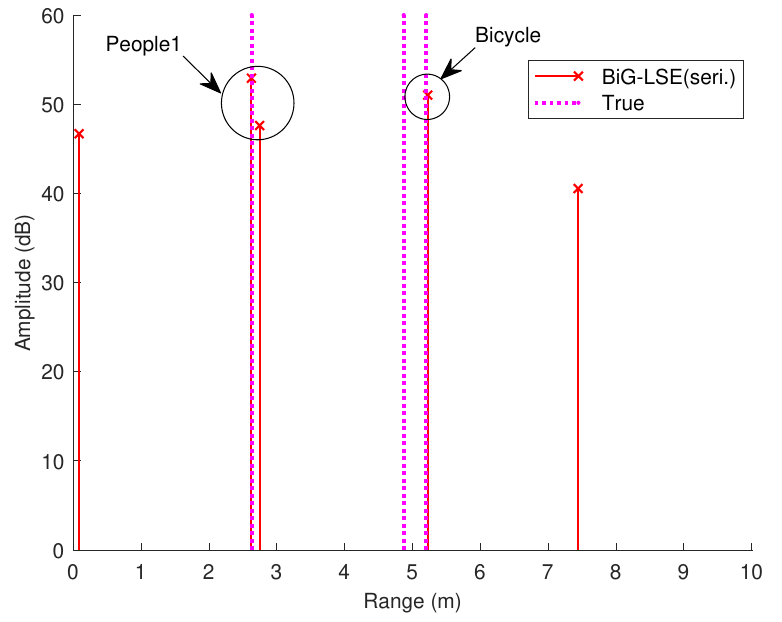}
		\label{exp_real_2_bl1}}
	\subfigure[]{
		\includegraphics[width=0.45\textwidth]{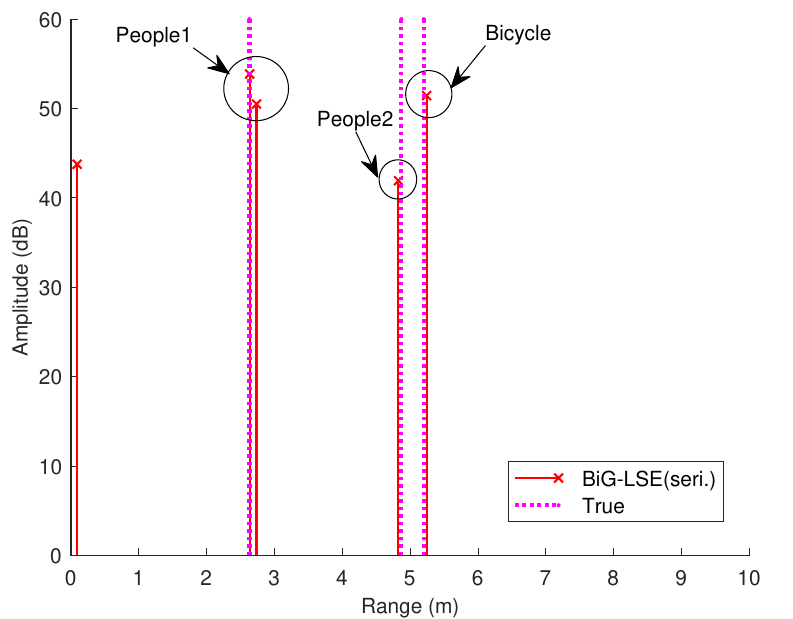}
		\label{exp_real_2_bl3}}
	\subfigure[]{
		\includegraphics[width=0.45\textwidth]{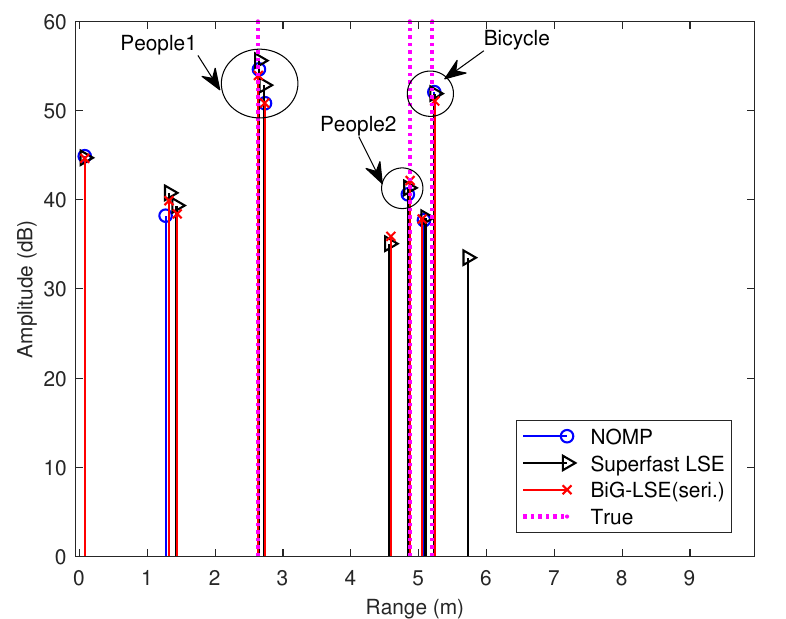}
		\label{exp_real_2_inf}}
	\caption{The range spectrum using FFT, the reconstruction result of the proposed algorithm under 1 bit quantization and 3 bit quantization: (a) range spectrum, (b) estimation results under 1 bit quantization, (c) estimation results under 3 bit quantization, (d) estimation results under 12 bit quantization.}
\end{figure*}

In this section, the performance of the proposed BiG-LSE (seri.) algorithm is demonstrated using real data acquired by the mmWave radar, including processing coarsely quantized data. The AWR1642 radar, an FMCW MIMO radar consisting of two transmitters and four receivers, is used to evaluate the algorithm's performance. The experimental setup is shown in Fig. \ref{exp_real_2}, and the parameters and waveform specifications are detailed in \cite{NOMPCFAR}. The radial distances of the two static individuals, referred to as People 1 and People 2, are approximately 4.87 m and 2.63 m, respectively. A cyclist moves toward the radar, starting at a radial distance of 7 m and approaching 2 m, with a velocity of about 2 m/s. The range spectrum, computed by the normalized DFT of the baseband data, is shown in Fig. \ref{exp_real_2_fft}. Only two peaks located at 2.75 m and 5.30 m are observed, corresponding to People 1 and the bicycle. 
The range estimation results for BiG-LSE (seri.) are obtained from 1-bit, 3-bit, and unquantized measurements, as shown in Figs. \ref{exp_real_2_bl1}, \ref{exp_real_2_bl3}, and \ref{exp_real_2_inf}, respectively. It is worth noting that the true noise variance is used in BiG-LSE (seri.) for the 1-bit quantization scenario, rather than the variance by EM learning. The noise variance can be computed as approximately 250 based on the data in Fig. \ref{exp_real_2_fft}. In the 3-bit quantization scenario, except for the energy leakage at 0 m, the algorithm successfully detects two individuals at 2.63 m and 4.83 m, and a moving bicycle at 5.24 m. In the 1-bit quantization scenario, the algorithm successfully detects People 1 at 2.62 m and the bicycle at 5.22 m but misses People 2 and generates a false target at 7.44 m. In the linear measurement scenario, all algorithms successfully detect People 1 at approximately 2.6 m, People 2 at approximately 4.8 m, and the bicycle at approximately 5.2 m, but also generate false targets at certain distances. Excluding the energy leakage at 0 m, Superfast LSE detects four false components at 1.32 m, 1.43 m, 4.57 m, and 5.72 m, BiG-LSE (seri.) detects three false components at 1.32 m, 1.44 m, and 4.59 m, while NOMP detects one false component at 1.27 m. It is observed that Superfast LSE generates the highest number of false targets, followed by BiG-LSE (seri.) and NOMP, which aligns with the simulation results.

\section{Conclusion}
In this paper, we propose BiG-LSE, derived using EP and approximation of the posterior PDFs for LSE. BiG-LSE automatically estimates the noise variance, the nuisance parameters of the prior distribution, provides uncertain degrees of the frequency estimates, and effectively handles nonlinear measurement scenarios. Two initialization schemes, based on frequency discretization and greedy initialization, are implemented, and variants of BiG-LSE incorporating the von Mises distribution for frequency estimation are developed. Extensive numerical simulations and a real experiment demonstrate that BiG-LSE with greedy initialization achieves high reconstruction accuracy while maintaining competitive runtimes, compared to state-of-the-art methods.

\bibliographystyle{IEEEbib}
\bibliography{strings,refs}

\end{document}